\documentclass[preprint]{elsarticle}

\usepackage[T1]{fontenc}
\usepackage{lineno,hyperref}
\usepackage{amsmath}
\usepackage{cleveref}
\usepackage{url}            
\usepackage{booktabs}       
\usepackage{amsfonts}       
\usepackage{nicefrac}       
\usepackage{microtype}      
\usepackage{graphicx}
\usepackage{braket}
\usepackage{macros}
\usepackage{subfigure} 
\usepackage{color}
\usepackage{comment}
\usepackage{float}
\usepackage{bbm}
\usepackage{geometry}

\usepackage{fancyhdr}
\fancypagestyle{firstpage}{%
        
        \lhead{}
        \rhead{\small
        TTK-24-07}
}

\newcommand{\ash}[1]{{\color{blue}{#1}}}

\journal{Physics Letters B}

\begin{document}

\begin{frontmatter}

\title{Machine learning mapping of lattice correlated data}

\author[1]{Jangho Kim}
\ead{j.kim@fz-juelich.de}

\author[2]{Giovanni Pederiva}
\ead{g.pederiva@fz-juelich.de}

\author[3,4,5]{Andrea Shindler}
\ead{shindler@physik.rwth-aachen.de}

\affiliation[1]{organization={Institute for Advanced Simulation (IAS-4) - Forschungszentrum J\"ulich}, 
addressline={Wilhelm-Johnen-Stra{\ss}e},
city={J\"ulich},
postcode={52428},
country={Germany}}

\affiliation[2]{organization={J\"ulich Supercomputing Centre (JSC) \& Center for Advanced Simulation and Analytics (CASA)}, 
addressline={Wilhelm-Johnen-Stra{\ss}e},
city={J\"ulich},
postcode={52428},
country={Germany}}

\affiliation[3]{organization={Institute for Theoretical Particle Physics and Cosmology, TTK, RWTH Aachen University},
addressline={Sommerfeldstr. 16},
city={Aachen},
postcode={52074},
country={Germany}}

\affiliation[4]{organization={Nuclear Science Division, Lawrence Berkeley National Laboratory},
city={Berkeley}, 
postcode={CA 94720},
country={USA}}%

\affiliation[5]{organization={Department of Physics, University of California},
city={Berkeley},
postcode={CA 94720},
country={USA}}%

\begin{abstract}
	We discuss a machine learning (ML) regression model to reduce 
	the computational cost of disconnected diagrams in lattice QCD calculations.
	This method creates a mapping between the results of fermionic loops 
	computed at different quark masses and flow times.
	The ML mapping, trained with just a small fraction of the complete data set,
  makes use of translational invariance
	and provides consistent result with comparable uncertainties over the calculation done over the whole ensemble, resulting in a significant computational gain.
\end{abstract}


\end{frontmatter}

\thispagestyle{firstpage}
\section{Introduction}
\label{sec:intro}

One of the computational challenges in lattice QCD calculations 
lies in the determination of the quark propagator, 
which not only serves as the foundation for calculating any fermionic 
correlation function but is also required in generating gauge ensembles 
with dynamical quarks. Computing the quark propagator involves inverting a 
very large sparse matrix representing the lattice Dirac operator. 
Fermionic disconnected diagrams appear in most hadron matrix element calculations,
as well as studies of flavor singlet channels, and standard methods for their calculation 
are based on stochastic estimates, which are usually computationally expensive. 

In this study, we aim to leverage on recent advancements in Machine Learning (ML) applications to lattice QCD calculations used to reconstruct the Euclidean time dependence of complex observables by correlating them with simpler functions~\cite{Yoon:2018krb}. The findings of Ref.~\cite{Yoon:2018krb} 
highlighted that ML techniques can effectively map various correlation functions, such as 2- and 3-point functions when utilizing the same Markov chain. Building upon this observation we extend this approach to calculate fermion disconnected diagrams. These calculations involve the manipulation of significant amounts of data, dependent on the amount of stochastic sources and gauge configurations employed.

Moreover, exploiting the inherent translational invariance of the lattice theory, we augment our dataset to thoroughly investigate correlations.
Utilizing numerous stochastic sources and translational invariance, we establish both training and bias-correction sets, thereby strengthening the robustness and precision of our analyses.

The gradient flow \cite{Luscher:2010iy,Luscher:2013cpa} provides a favorable 
regulator of short-distance singularities 
due to its reduced operator mixing, essentially trading power divergent lattice spacing 
effects with a milder finite $1/t$ dependence. 
By keeping the flow time $t$ fixed, one can then perform the
continuum limit with no renormalization ambiguities. 
An example of the advantage of the use of the Gradient Flow is the 
simplified calculation of the quark content of nucleons~\cite{Shindler:2014oha,Shindler:2023xpd} or the resolution of the problem of power 
divergences for higher dimensional operators~\cite{Kim:2021qae}.
The application of the gradient flow to the calculation of fermionic disconnected diagrams is beneficial both to simplify the renormalization and to improve the signal-to-noise ratio.

In Sec.~\ref{sec:disc} we describe the stochastic method we use to determine
the fermionic disconnected diagrams. In Sec.~\ref{sec:corr} we study 
the correlation between data and, in Sec.~\ref{sec:ml} we describe the algorithm
and present our results.

\section{Fermionic disconnected diagrams}
\label{sec:disc}

We consider a lattice of spacing $a$ and box size $V = L^3 \times T$,
with Dirac fermions in the fundamental representation of 
SU($\Nc = 3$),
$\psi_{a,\alpha},\psibar_{a,\alpha}$ ($a=1,\ldots,\Nc = 3$ and $\alpha=1,\ldots,4$).
We adopt periodic boundary conditions for all fields, with the exception that the boundary condition in Euclidean time is anti-periodic for fermion fields.
These boundary conditions preserve translational invariance.
In lattice QCD, the calculation of physical observables involving fermions,
requires the determination of the quark propagators,
$\left[ \psi_{a,\alpha}(x)\psibar_{b,\beta}(y)\right]_{\text{F}} = 
S^{ab}_{\alpha\beta}(x,y)$, where with $\left[ \cdot \right]_{\text{F}}$ we indicate a fermion contraction.
For simple correlation functions,
like the kaon or the nucleon 2-point functions, 
one requires only the calculation of one column of the inverse, $S$, of the lattice 
Dirac operator, $D$,
\be 
D^{ab}_{\alpha\beta}(x,y)S^{bc_0}_{\beta\gamma_0}(y,z_0)=
\frac{1}{a^4}\delta_{ac_0}\delta_{\alpha\gamma_0}\delta_{x z_0}\,, 
\label{eq:dirac_eq}
\ee 
where the source location, $z_0$, and the corresponding spin and color indices, $\gamma_0$ and $c_0$,
are fixed. Repeated indices, $\beta$, $b$, and $y$ in this case, are summed over.
To determine all-to-all propagators, where the source location is not fixed, or
to calculate quantities related to all-to-all propagators, like the trace of the quark 
propagator, we have to rely on stochastic methods.

To calculate the all-to-all propagator one takes
a set of $r= 1, \ldots, N_\eta$ of complex random vectors
$\eta^{(r)}_{a\alpha}(x)$ that satisfy
\be 
\lim_{N_\eta \rightarrow \infty} \llangle \eta_{a\alpha}(x)\eta_{b\beta}(y)^* \reta = 
\delta_{ab}\delta_{\alpha\beta}\delta_{xy}\,, 
\ee
where with $\llangle \cdot \reta$ we indicate the average over
$N_\eta$ stochastic vectors
\be 
\llangle \eta_{a\alpha}(x) \eta_{b\beta}(y)^* \reta = 
\frac{1}{N_\eta} \sum_{r=1}^{N_\eta} \eta_{a \alpha}^{(r)}(x) \eta_{b \beta}^{(r)*}(y)\,.
\label{eq:eta_delta}
\ee  
To estimate the all-to-all propagator one can now solve
\be 
D^{ab}_{\alpha\beta}(x,y)\phi_{b\beta}(y)=\frac{1}{a^4}\eta_{a\alpha}(x)\,.
\label{eq:dirac_stochsource}
\ee 
The full propagator is then reconstructed by the unbiased estimator 
\be 
S^{ab}_{\alpha\beta}(x,y) = 
\lim_{N_\eta \rightarrow \infty} \left\langle \phi_{a \alpha}(x)\eta_{b \beta}(y)^* \reta\,,
\ee 
up to noise contributions at finite $N_\eta$.
The relative total noise of the estimator is of the order of O($\sqrt{12 V /N_\eta}$) and one needs variance reduction techniques to reach a signal-to-noise ratio of O($1$).
One example of variance reduction would be the use of 
time-dilution~\cite{Foley:2005ac}.
Another choice is to use the one-end trick~\cite{Foster:1998vw} and the generalization called 
{\it{linked}} stochastic sources~\cite{ETM:2008zte} where the stochastic vector 
is non-vanishing only for specific color, spin or space-time indices.
We denote linked stochastic sources with 
$\eta^{(b,\beta)}_{a \alpha}(x)$ where the color and Dirac indices $b$ and $\beta$
are fixed
\be  
\eta^{(b,\beta)}_{a \alpha}(x) = 
\delta_{ab}\delta_{\alpha \beta}\eta_{a\alpha}(x)\,.
\label{eq:linked}
\ee 
We now solve for all the fixed couple of values $(a_0, \alpha_0)$ 
\be 
D^{ab}_{\alpha\beta}(x,y)\phi^{(a_0 \alpha_0)}_{b \beta}(y)=
\frac{1}{a^4}\eta^{(a_0 \alpha_0)}_{a\alpha}(x)\,.
\label{eq:dirac_stochsource2}
\ee
For different quark flavors, $f=\ell,s$, one obtains different solutions $\phi^f(y)$.
To not clutter the notation, we leave the flavor index unspecified 
when discussing the generalities of the quark propagator determination.
The quantity of interest is the quark propagator, $S^{a_0 b_0}_{\alpha_0 \beta_0}(x,y)$ that can 
be determined for each gauge configuration up to noise contributions by 
\be 
\left[\psi_{a_0,\alpha_0}(x) \psibar_{b_0,\beta_0}(y)\right]_{\text{F}} = 
S^{a_0 b_0}_{\alpha_0 \beta_0}(x,y) =
\llangle  \phi^{(a_0,\alpha_0)}_{c \gamma}(y) \eta^{(b_0 \beta_0)}_{c \gamma}(x)^*
\reta\,.
\label{eq:prop_linked}
\ee 

Reintroducing a flavor index, the quark condensates at vanishing flow times 
is denoted by 
\be 
\left\langle \psibar_f \psi_f \right\rangle = 
\frac{a}{T} \sum_{x_4} \left\langle C^f_{\psibar \psi}(x_4;\eta,U) \right\rangle_{\text{G},\eta}\,, 
\ee 
where $\left\langle \cdot \right\rangle_{\text{G},\eta}$ denotes 
the average over the gauge ensemble and the stochastic sources.
$C^f_{\psibar \psi}(x_4;\eta,U)$ is evaluated on a
fixed gauge background, $U(x,\mu)$, and on a given stochastic source, $\eta(x)$ 
\be
C^f_{\psibar \psi}(x_4;\eta,U) = - \frac{a^3}{L^3} \sum_{\bx} \sum_{a_0,\alpha_0} 
\phi^{f,(a_0 \alpha_0)}_{c \gamma}(\bx,x_4)\eta^{(a_0,\alpha_0)}_{c \gamma}(\bx,x_4)^* \,,\qquad f=\ell,~s\,.
\label{eq:chiral_time}
\ee
The specific choice of the stochastic vector $\eta(x)$ is not critical as far as the condition \eqref{eq:eta_delta} is satisfied and its variance remains within acceptable limits.
For a complex matrix like $D^{ab}_{\alpha\beta}(x,y)$ 
a standard choice is to use stochastic vectors 
belonging to $\mathbbm{Z}_4$, i.e for each $a,\alpha,x$ the vector 
$\eta_{a\alpha}(x)$ takes one of the values 
$\{\pm 1, \pm i\}$ with the same probability 
(see Ref.~\cite{HeatherM:2021gbh} and refs. therein for a discussion on
the choice of stochastic vectors and variance reduction techniques).

In this work we also consider the flowed scalar quark condensate 
\be 
\left\langle \chibar_f \chi_f \right\rangle = 
\frac{a}{T} \sum_{x_4} \left\langle C^f_{\chibar \chi}(x_4,t;\eta,U) \right\rangle_{\text{G},\eta}
\ee 
where 
\be
C_{\chibar \chi}(x_4,t;\eta,U) = \frac{a^3}{L^3} \sum_{\bx} \sum_{a,\alpha} 
\left[\chibar^a_{\alpha}(\bx,x_4,t)\chi^a_{\alpha}(\bx,x_4,t)\right]_{\text{F}}\,.
\label{eq:cond_flow}
\ee 
It is assumed that the fermion fields $\chi(x,t)$ and $\chibar(x,t)$
satisfy the gradient flow equations of Refs.~\cite{Luscher:2010iy,Luscher:2013cpa},
but the results of this work do not depend on the 
particular choice of gradient flow equations.

The first step is to solve for each pair $(a_0, \alpha_0)$ the equation
\be 
D^{ab}_{\alpha\beta}(x,y)\phi^{(a_0 \alpha_0)}_{b \beta}(y;0,t)=
\frac{1}{a^4}\xi^{(a_0 \alpha_0)}_{a\alpha}(x;t,0)\,,
\label{eq:dirac_stochsource_flow}
\ee
where the source
\be 
\xi^{(a_0 \alpha_0)}_{a\alpha}(x;t,0) = a^4 \sum_u K(u,x;t,0)^\dagger \eta^{(a_0 \alpha_0)}_{a\alpha}(u)
\ee
has to be determined for each value of the flow time $t$
solving the adjoint flow equation
\be 
\left(\partial_s + \Delta\right)\xi^{(a_0 \alpha_0)}_{a\alpha}(x;t,s) = 0\,, 
\quad \xi^{(a_0 \alpha_0)}_{a\alpha}(x;t,t) = \eta^{(a_0 \alpha_0)}_{a\alpha}(x)\,,
\label{eq:adjoint_flow}
\ee 
to $s=0$.
The kernel $K(x,y;t,s)$, defined in Ref.~\cite{Luscher:2013cpa}, is the solution of the gradient flow equation 
\be 
\left(\partial_t - \Delta \right)K(x,y;t,s) = 0\,, \quad 0 \le s \le t \,,
\ee
with $K(x,y;t,t) = 1/a^4 \delta_{xy}$ for all $t\ge 0$.
The stochastic vector $\eta^{(a_0 \alpha_0)}_{a\alpha}(x)$ is the linked vector adopted 
in the $t=0$ case defined in Eq.~\eqref{eq:linked}.
The flowed scalar condensate can then be determined on a single gauge configuration by the 
expression 
\be
C^f_{\chibar \chi}(x_4,t;\eta,U) = - \frac{a^3}{L^3} \sum_{\bx} \sum_{a_0,\alpha_0} a^4 \sum_y
K(x,y;t,0) \phi^{(a_0 \alpha_0)}_{c \gamma}(y;0,t) 
\eta^{(a_0,\alpha_0)}_{c \gamma}(x)^*\,,
\label{eq:cond_time_flow}
\ee
where to compute 
\be 
\phi^{(a_0 \alpha_0)}_{c \gamma}(x;t,t)  = 
a^4 \sum_y K(x,y;t,0) \phi^{(a_0 \alpha_0)}_{c \gamma}(y;0,t)\,,
\ee 
is sufficient to solve the gradient flow equation  
\be 
\left(\partial_s - \Delta \right)\phi^{(a_0 \alpha_0)}_{c \gamma}(x;s,t) = 0\,,
\ee 
for $s=t$ where the initial condition at $s=0$ is given by the solution of Eq.~\eqref{eq:dirac_stochsource_flow}, $\phi^{(a_0 \alpha_0)}_{c \gamma}(y;0,t)$.

\section{Correlation maps and translation invariance}
\label{sec:corr}

\begin{figure}[h]
{
 \includegraphics[width=0.45\textwidth]{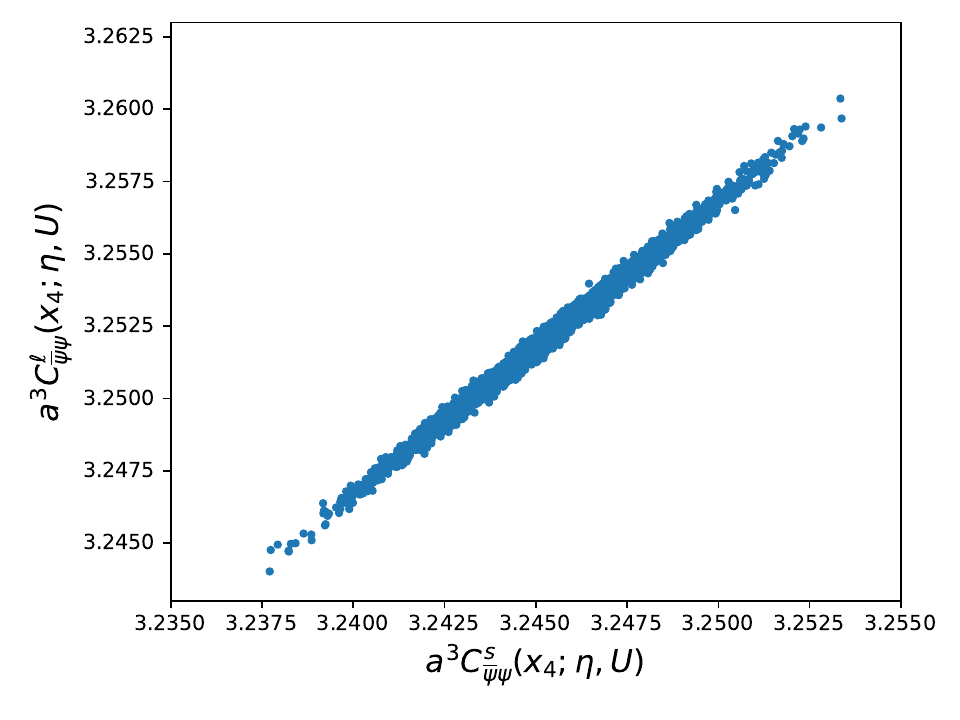}
 }
{
 \includegraphics[width=0.45\textwidth]{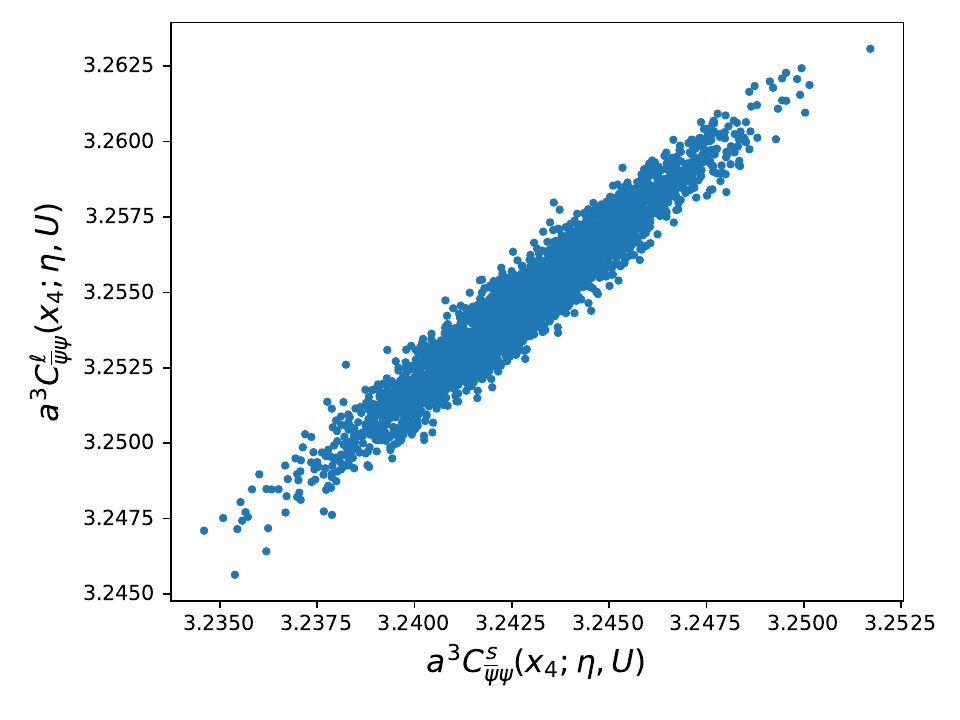}
	 }
	 \caption{
    Graphical representation of the correlation
    between the light and strange quark scalar condensates at $t=0$ for a single stochastic source for $M_1$ and $M_3$ ensembles, $N_{G,T} = 50$ configurations and $T/a=64$ Euclidean time coordinates. These are the sets that have been used to train the ML mapping (see Sec.~\ref{sec:ml}).	 \label{fig:corr_train}}
\end{figure}

For our numerical experiment we consider the lattice ensembles 
summarized in Table~\ref{tab:ensembles}.
\begin{table}[ht]
  \centering
  \begin{tabular}{|c|c|c|c|c|c|c|c|c|c|c|c|c|}
    \hline
		Label & $\beta$ & $\kappa_\ell$ & $\kappa_s$ & $m_{\pi}[\text{MeV}]$ & $m_{K}[\text{MeV}]$ & $a[\text{fm}]$ & $L/a$ & $T/a$ & $\csw$ & $N_G$ \\
    \hline\hline
		M$_1$ & 1.90 & 0.13700 & 0.1364 & 699.0 & 789.0 & 0.0907 & 32 & 64 & 1.715 & 399  \\
    \hline                                                                         
		M$_3$ & 1.90 & 0.13754 & 0.1364 & 409.7 & 644.0 & 0.0907 & 32 & 64 & 1.715 & 450  \\
    \hline
  \end{tabular}
  \caption{Summary of the lattice bare parameters for the ensembles used in this work.
    $N_G$ is the number of gauge configurations selected from Ref.~\cite{Aoki:2010wm,PACS-CS:2008bkb}.
		All the other labels should be self-explanatory. 
    \label{tab:ensembles}}
\end{table}
They have been generated~\cite{Aoki:2010wm,PACS-CS:2008bkb} using $N_f=2+1$ dynamical fermion flavors all 
regulated with a non-perturbative O($a$) clover-improved lattice fermion action
and the Iwasaki gauge action.
\begin{figure}[t]
    \centering
   \subfigure[Euclidean time invariance]{	\label{fig:indep_Eucl} 
 \includegraphics[width=0.45\textwidth]{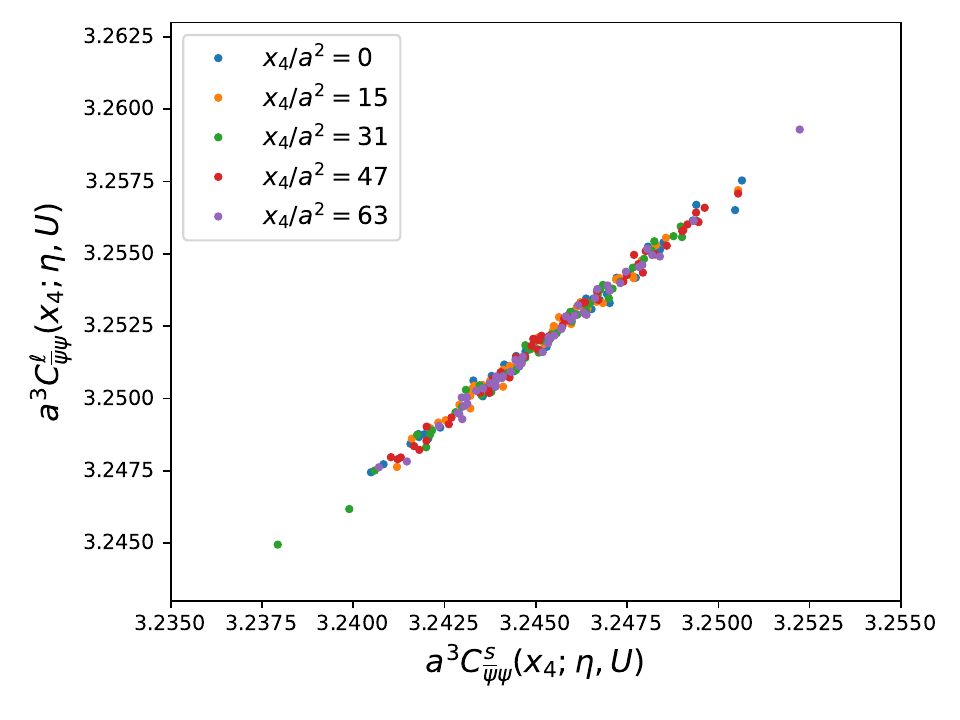}
   }
   \subfigure[Noise source independence]{	\label{fig:indep_src} 
 \includegraphics[width=0.45\textwidth]{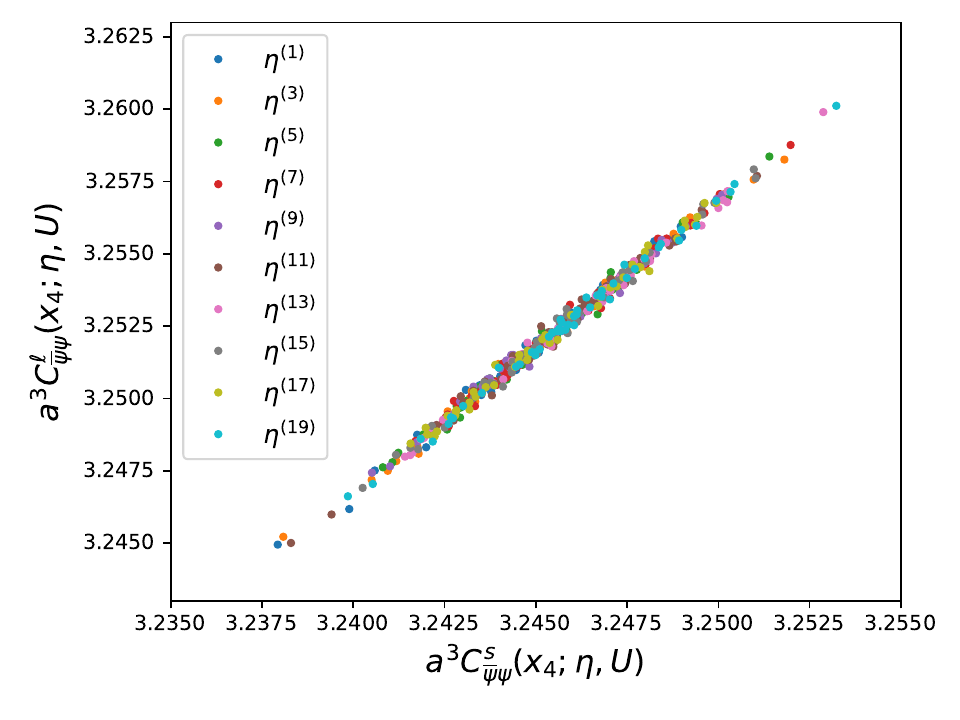}
   }
 \caption{Graphical representation of the correlation
 between the light and strange quark scalar condensates at $t=0$ for several gauge configurations,
 Euclidean time coordinates and stochastic sources on $M_1$ ensemble.
	Left: correlation between light and strange quark scalar condensates for selected values of $x_4$.
  The data shown correspond to the training set of $N_{G,T} = 50$ gauge configurations 
  and $N_\eta = 1$ stochastic source.
  Right: Same as the left plot. The data shown correspond to $N_{G,T} = 50$, a fixed value of $x_4 = T/2$ 
  and a selection of stochastic sources.
			 \label{fig:corr}
			 } 
\end{figure}
On these ensembles we have calculated, using stochastic sources 
as described in the previous section, the Euclidean time 
dependence of the unflowed and flowed light and strange scalar condensates.
On a fixed background gauge configurations, they are determined using Eqs.~\eqref{eq:chiral_time} and \eqref{eq:cond_time_flow}, respectively.
We have then analyzed the correlation between the $2$ observables.
In Figs.~\ref{fig:corr}- \ref{fig:corr_train} we show correlation plots between the light and strange quark condensates at vanishing flow time, $t=0$, calculated on $M_1$ and $M_3$ ensembles. 
We observe a strong correlation of the data independently on the 
Euclidean time where we calculate the condensate (see Fig.~\ref{fig:indep_Eucl}).
This is a consequence of translational invariance and is consistent with the observation 
that averaging over all lattice points provides a better statistical precision,
making use of the full lattice. In this context we want to take advantage 
of translational invariance to enlarge the data set 
used to train the ML mappings.
A similar strong correlation is observed varying the stochastic source used 
for the determination of the quark propagator (see Fig.~\ref{fig:indep_src}).
This observation enables us to use one or more stochastic sources in the training set, allowing for a relatively small number of gauge configurations for training. In other words, this approach increases the dimensionality of the data space that is partitioned into training, bias, and unlabeled sets.
In Fig.~\ref{fig:corr_train} we show the correlation plot used to train the ML model 
on the ensembles $M_1$ and $M_3$, where we have used $N_{G,T} = 50$ gauge configurations, all the
$64$ Euclidean time values of $x_4/a$, and $N_{\eta}=1$ stochastic source.
Details on the choice of the training set are discussed in Sec.~\ref{sec:ml}.
From Fig.~\ref{fig:corr_train} we note that 
for ensembles at lighter pion masses 
a similar, but slightly weaker, correlation is measured.
The weaker correlation could be caused by a generic loss of correlation
for lighter pion ensembles or by a larger mass difference between 
the correlated observables.
This does not prevent us to successfully test our ML method also 
for the ensemble $M_3$.

We have observed a similar strong correlation 
also between condensates calculated at different 
flow times on the same ensembles.
The expected correlation between observables evaluated on the same ensembles in the context of ML modeling could help speed up the calculation of observables from lattice QCD simulations.

\section{Decision tree mapping of correlations}
\label{sec:ml}

To take advantage of the correlations observed and described in the 
previous section we have scrutinized a few supervised machine learning 
methods, and we have found, like it was found in Ref.~\cite{Yoon:2018krb}, that a decision tree (DT) is sufficient to capture the correlation between data.
There are perhaps other maps, like specific neural networks,
that might be able to describe the correlations equally well or even better 
in certain cases, but for this first investigation DT is sufficiently accurate.

\subsection{Description of the algorithm}

Decision Tree (DT) stands as a non-parametric supervised 
learning technique employed for regression tasks involving
the prediction of continuous numerical outcomes, by recursively partitioning 
the input space into subsets based on feature conditions 
and assigning a constant value to each resulting 
region~\cite{breiman1984classification,breiman2017classification}. 
The primary aim is to build a model that can predict the 
value of a target variable by acquiring basic decision rules 
deduced from the features of the data.

Based on the observation~\cite{Yoon:2018krb} that DTs can map correlations between lattice QCD data, we trained a DT to determine
fermion disconnected diagrams at a given quark mass, or flow time,
given a fermion disconnected diagram calculated at a different quark mass 
or flow time. 
We take a subset of the total amount of data to train a ML model, 
a DT in this case, and calculate the corresponding bias, taking advantage 
of the many stochastic sources used for the calculation of disconnected diagrams and translational invariance of the lattice theory.
To train the DT we divide the total set of data $N$ into labeled, $N_L$, 
and unlabeled data, $N_U$.
The labeled data are divided into the subset $N_T$ to train the 
machine learning (ML) model, 
and the subset $N_B$ to estimate the bias correction, with $N_L = N_T + N_B$.
The data we consider in this numerical experiment are 
fermionic disconnected diagrams calculated at $2$ different quark masses
or flow times.
If we denote with $N_\eta$ the number of stochastic sources, 
$N_G$ the number of gauge configurations,
and we make use of translational invariance, we can use 
as a complete set of data for a given condensate
$N = N_\eta \times N_G \times T/a$ points.
As labeled data we consider the subset constituted by 
$N_L = N_{\eta,L} \times N_G \times T/a$,
further divided into the training set $N_T = N_{\eta,L} \times N_{G,T} \times T/a$
and the bias correction set $N_B = N_{\eta,L} \times N_{G,B} \times T/a$, where 
$N_G = N_{G,T} + N_{G,B}$. We have decided to use the same number of sources for the training and bias set and utilize the full ensemble for the labeled and unlabeled data. Different choices are indeed possible and in Secs.~\ref{ssec:robust} and \ref{sec:conclusions} we analyze the dependence on the size of the training and bias set.
Once the ML model has been trained, it is applied to the unlabeled data 
$N_U = N_{\eta ,U} \times N_G \times T/a$, where $N_{\eta ,U} = (N_\eta - N_{\eta,L})$.
This implies that, fixed the labeled data, 
we have a single DT model for each pair of observables and each ensemble.

Unlike Ref.~\cite{Yoon:2018krb}, we partitioned the labeled and unlabeled datasets based on the selected stochastic sources, utilizing the entire ensemble for both. Additionally, we leveraged translation invariance and trained the ML model using features across the entire Euclidean time extent of the lattice. In alignment with the approach 
employed in Ref.~\cite{Yoon:2018krb}, we segmented the bias and training sets by stratifying the gauge configurations.

To illustrate, we consider determining an observable $O$
for given values of external parameters such as the quark mass 
$m$ or the flow time $t$. 
The DT is trained using features comprising the same observable, $O_f$,
at different values of the quark mass or flow time. These features
are evaluated on a subset of the total dataset, referred to as the training set.
The mapping obtained through this training between the two observables is denoted as $\Gamma_f$, where the subscript $f$ indicates that the mapping depends on the choice of the external parameters. In Sec.~\ref{ssec:robust}, we demonstrate that this mapping is independent of the choice and size of the training set.
\begin{figure}[t]
  \centering
   \subfigure[Distributions of the light quark condensate for the ensemble $M_1$ at $t/a^2=0$. The ML mapping has been trained with the strange quark condensate.]{	
  \includegraphics[width=0.30\textwidth]{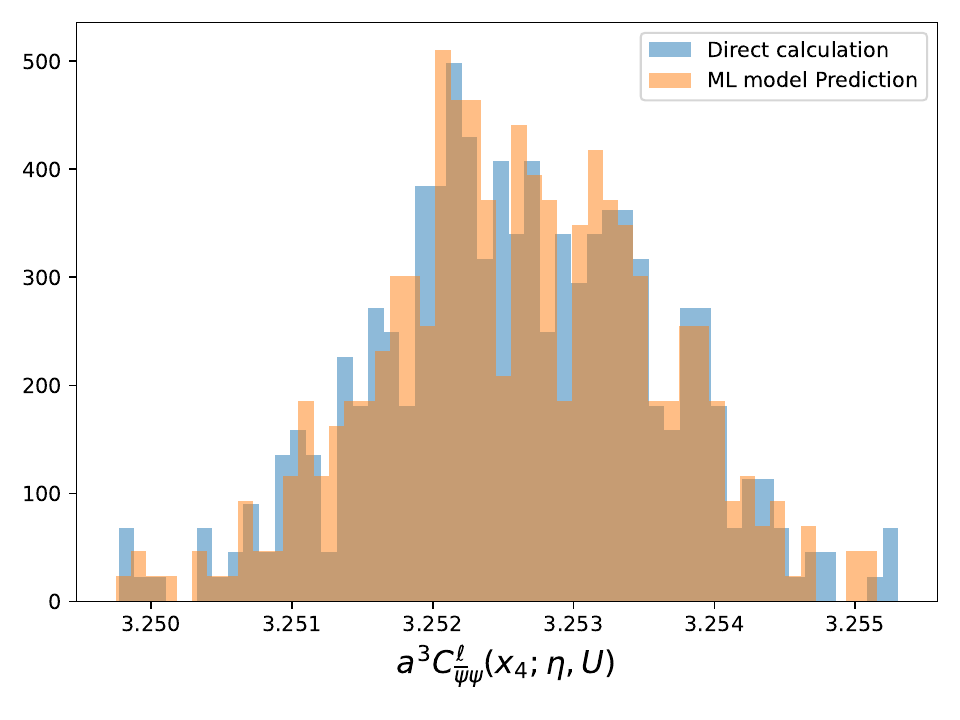}
   }
	 \hspace{0.1cm}
   \subfigure[Distributions of the light quark condensate for the ensemble $M_1$ at $t/a^2=0.7$. The ML mapping has been trained at $\bar{t}/a^2=0.5$.]{
         \includegraphics[width=0.30\textwidth]{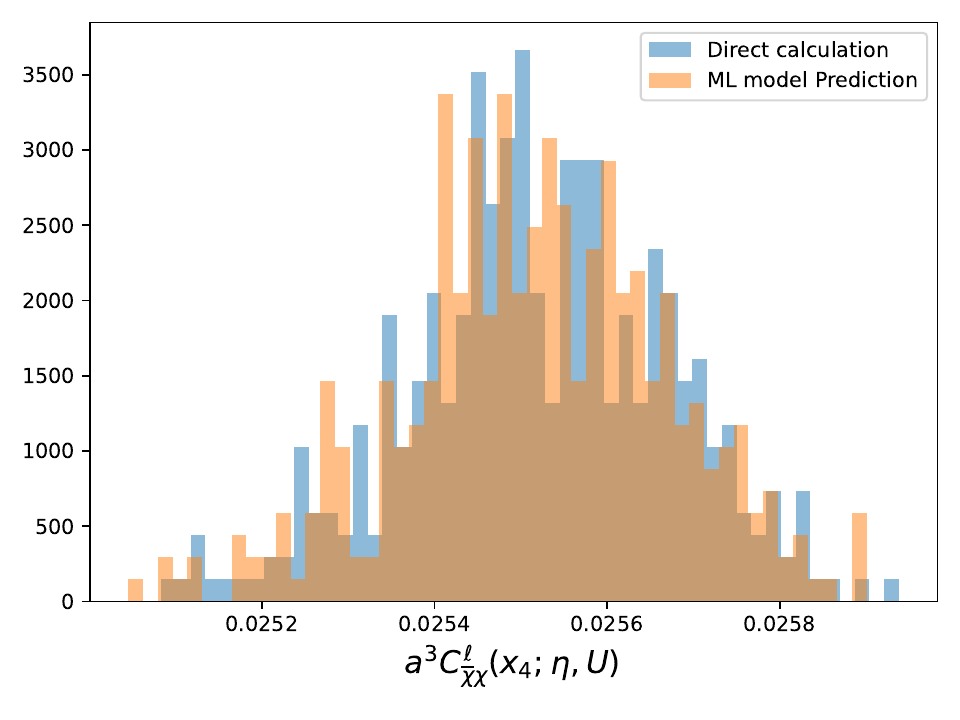}
   }
	 \hspace{0.1cm}
   \subfigure[Distributions of the light quark condensate for the ensemble $M_3$ at $t/a^2=0$. The ML mapping has been trained with the strange quark condensate.]{
         \includegraphics[width=0.30\textwidth]{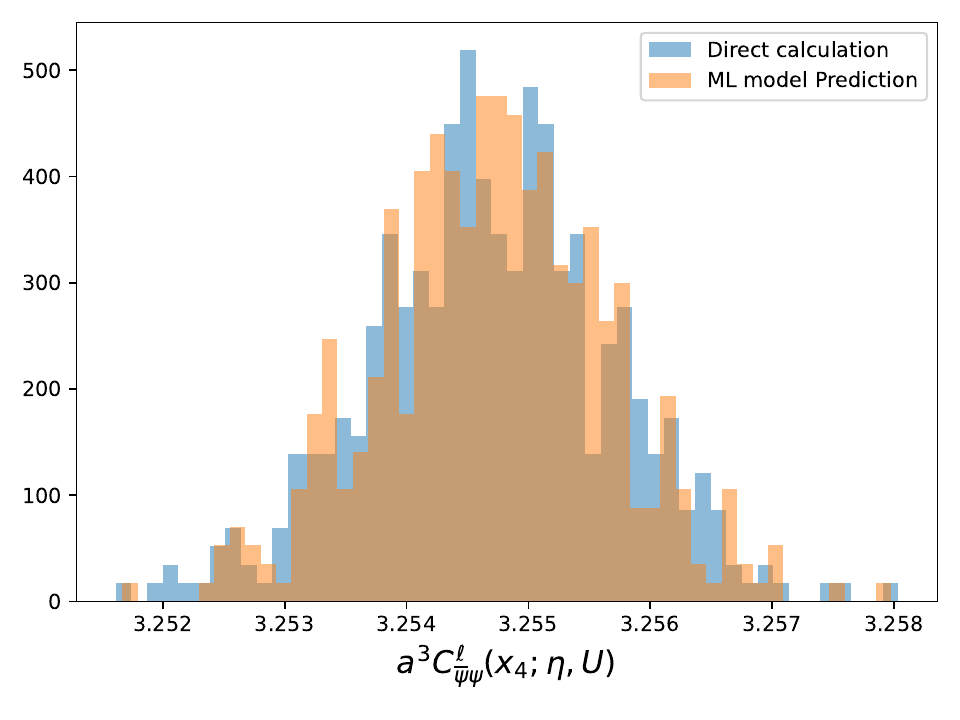}
   }
 \caption{
       \label{fig:corr_dt} Distributions on the unlabeled data of quark condensates at $x_4=T/2$ from a direct 
       calculation and from the ML mapping.
   }
\end{figure}
After the training, the target 
quantity $O_\Gamma(U,\eta,x_4)$ is obtained by applying the ML mapping to the 
features on the unlabeled data, i.e.
\be 
O_\Gamma(U,\eta,x_4) = \Gamma_f[O_f(U,\eta,x_4)]\,.
\ee 
The dependence of $O_\Gamma$ on the external parameters $m$, and $t$ is left implicit.
We keep explicit the dependence on the variables $U,\eta,x_4$ labelling 
the data set.
The ML mapping $\Gamma_f$ does not depend explicitly on the 
training data, but it is evaluated on features that 
depend on the specific gauge configuration, stochastic source, and 
Euclidean time and thus so is the output observable.

The correlation function is then obtained averaging on the unlabeled data 
\be  
\frac{1}{N_G} \sum_U \frac{1}{N_{\eta ,U}} \sum_{\eta_U}O_\Gamma(U,\eta_U,x_4)\,,
\label{eq:corr_ML}
\ee 
where with $\sum_{U}$ we indicate a sum of the gauge configurations and $\sum_{\eta_U}$
indicates the sum over all the sources belonging to the unlabeled data.
In Sec.~\ref{ssec:bias} we discuss the results including a bias correction.

The first example we consider is 
$O_f=C^s_{\psibar \psi}(x_4;\eta,U)$,
i.e. the strange scalar quark condensate, 
and  $O = C^l_{\psibar \psi}(x_4;\eta,U)$, i.e. the light scalar quark condensate.
Both are defined in Eq.~\eqref{eq:chiral_time}.
The second example we consider is $O_f=C^l_{\chibar \chi}(x_4,\bar{t};\eta,U)$
i.e. the light quark scalar condensate at flow time $\bar{t}$ , and  
$O= C^l_{\chibar \chi}(x_4,t;\eta,U)$, 
i.e. the same condensate at a different flow time $t$. 
This quantity is defined in Eq.~\eqref{eq:cond_time_flow}.

The DT mapping, $\Gamma$, is determined on the training set minimizing 
a loss function given by the mean squared error,
and for each node, the algorithm considers all the 
input data and chooses the best split; nodes are expanded until all leaves
contain a single sample.
A brief discussion and the summary of the other hyperparameters of the model
is found in \ref{app:hyperparams}.

The mapping is then applied to the unlabeled data 
of the input quantity for each $N_{\eta,U} = N_\eta - N_{\eta,L}$ 
stochastic source, each gauge configuration, and each $x_4/a$.
To avoid autocorrelation in the training procedure,
we have chosen the set of $N_{G,T}=50$ gauge configurations, each maximally 
separated in the Markov chain. Specifically, we select gauge configurations 
separated by $40$ and $45$ molecular dynamics trajectories for the two ensembles.
When applying the ML mapping to the unlabeled data, we first average over the 
stochastic sources, then build blocks of $7$ elements for the ensemble $M_1$ and $9$ elements for the ensemble $M_3$. Subsequently,  
a standard bootstrap procedure is applied with $N_b=1000$ bootstrap samples.
This allows for the calculation of the statistical error of the resulting condensate.

We have studied the dependence of the signal-to-noise ratio (SNR) of the scalar
condensate on the number of stochastic sources, $N_\eta$, for $t/a^2=0$ and 
$t/a^2=0.5,~0.7$ as representative positive flow times, and the full set 
of gauge configurations available in the ensembles 
labeled $M_1$ and $M_3$
(see Table~\ref{tab:ensembles} in the next section). 

We did not attempt to optimize the number of sources and chose 
$N_\eta = 20$ for all our calculations.\footnote{
  For the analysis at flow time $t/a^2=1.0$ for ensemble $M_3$, 
  we used a total of $N_\eta=30$ stochastic sources (see Table~\ref{tab:gain}).}

Even though this might not be the optimal choice,
it is sufficient to test 
the ML algorithm we propose in this work.
It is worth noting that our numerical experiments indicate
that a different number of stochastic sources would be needed to saturate the SNR for $t>0$,
as a result of the smoothing effect of the gradient flow.

   For the training we use $N_{\eta,L}=1$ stochastic source 
and $N_{G,T}=50$ gauge configurations. 
This gives us, for both ensembles, 
a total of $N_T=3200$ data points to train the ML model.

In Fig.~\ref{fig:corr_dt}, we present the distributions of three condensates, comparing the results obtained from the direct calculation of the condensates with those derived from the ML mapping. This comparison illustrates 
how the ML approach can replicate the result obtained 
with a direct calculation capturing the essential features of the distribution of the condensates. For a more quantitative analysis, we present averages and statistical errors in the next section, including bias corrections for further accuracy.

\subsection{Results}
\label{ssec:bias}

The results obtained with the ML mapping, cfr. Eq.~\eqref{eq:corr_ML},
have been corrected for possible biases. 
To estimate the bias correction we use the same stochastic sources 
chosen for the training set, $\eta_L$, and the remaining 
$N_{G,B} = N_G - N_{G,T}$ gauge configurations. 
The bias-corrected result is given by 
\bea
\left\langle O_\Gamma \right\rangle_{G,\eta}(m,t,x_4) &=& 
\frac{1}{N_G} \sum_U \frac{1}{N_{\eta ,U}} \sum_{\eta_U}O_\Gamma(U,\eta_U,x_4) + \nonumber \\
&+& \frac{1}{N_{G,B}} \sum_{U_B} \frac{1}{N_{\eta ,L}} \sum_{\eta_L}
\left[O(U_B,\eta_L,x_4) - O_\Gamma(U_B,\eta_L,x_4)\right]\,,
\label{eq:bias_corr}
\eea
where the bias correction is evaluated on the bias set.
The calculation of the bias presented in this section is an extension of the 
calculation of Ref.~\cite{Yoon:2018krb}, where we include the possibility of choosing different stochastic sources where the fermionic disconnected diagram is evaluated.

\begin{figure}[t]
  \subfigure[Without bias correction]{
 \label{fig:pred_M1_flavor_a}
  \includegraphics[width=0.3\textwidth]{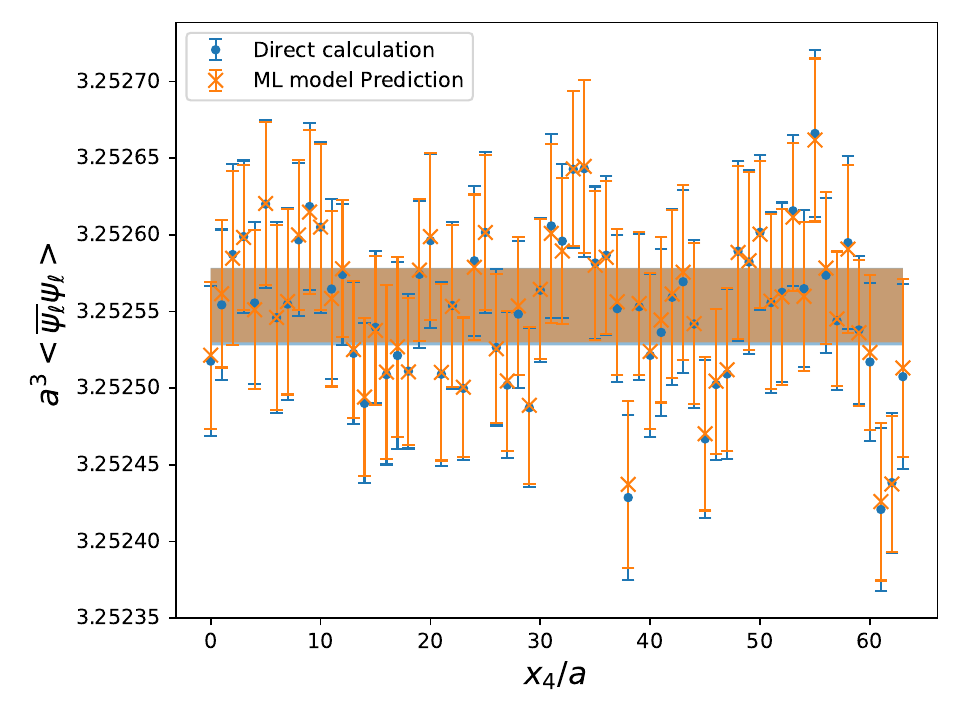}
  }
	\subfigure[With bias correction]{
	\label{fig:pred_M1_flavor_b}
  \includegraphics[width=0.3\textwidth]{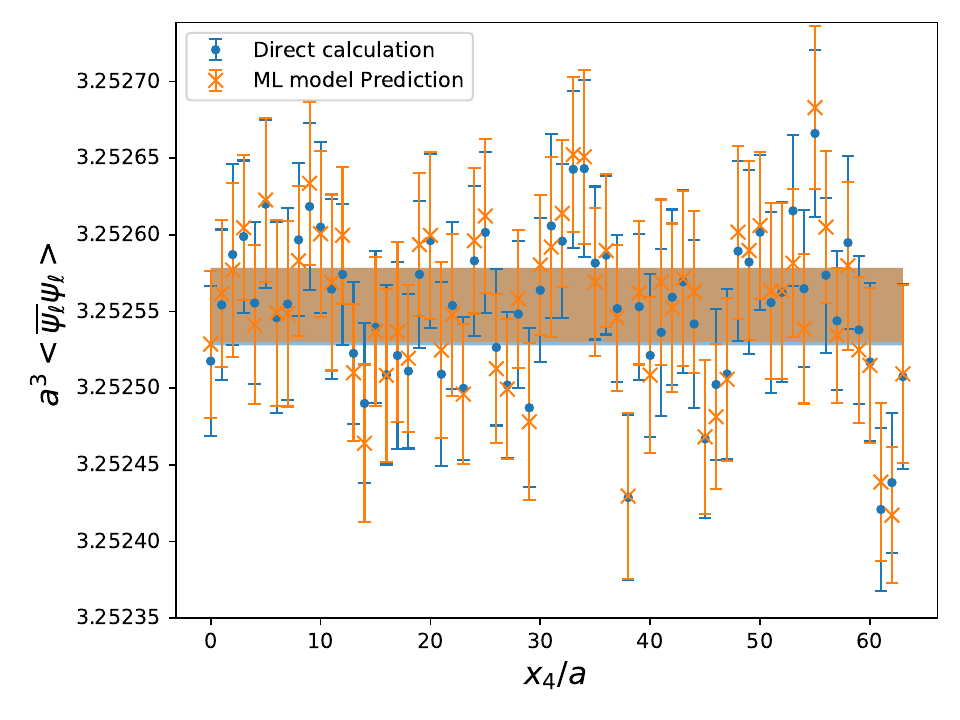}
  }
  \subfigure[Deviation between the direct and the ML calculation 
  normalized with the total standard deviation (see main text)]{ 
 \label{fig:pred_M1_q}
  \includegraphics[width=0.3\textwidth]{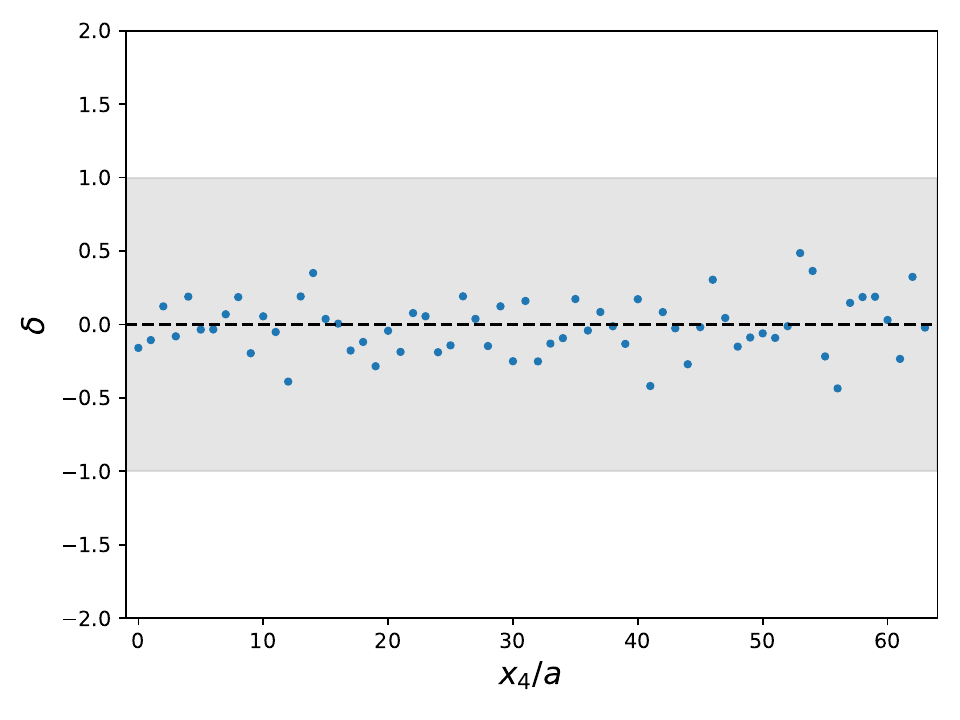}
  }
  \caption{ \label{fig:pred_M1_flavor} 
  Comparison of the light quark condensate on the ensemble $M_1$ between the ML result and the
  direct (standard) calculation of the same condensate as a function of Euclidean time. 
  The comparison is performed on the unlabeled data.
  }
\end{figure}
We have determined the light quark condensate 
at $t=0$ training the ML mapping as described in the previous section
using the strange quark condensate as features, and then calculated the bias corrections as described above.
We have also used a ML mapping to determine the flowed scalar quark condensate
on a set of flow times. 
The features used to train the ML mapping are the light quark condensate at $\bar{t}/a^2=0.5$ 
for the ensemble $M_1$ and $\bar{t}/a^2=1.0$ for the ensemble $M_3$.

\begin{figure}[t]
  \subfigure[Without bias correction]{
 \label{fig:pred_M1_flowtime_a}
  \includegraphics[width=0.3\textwidth]{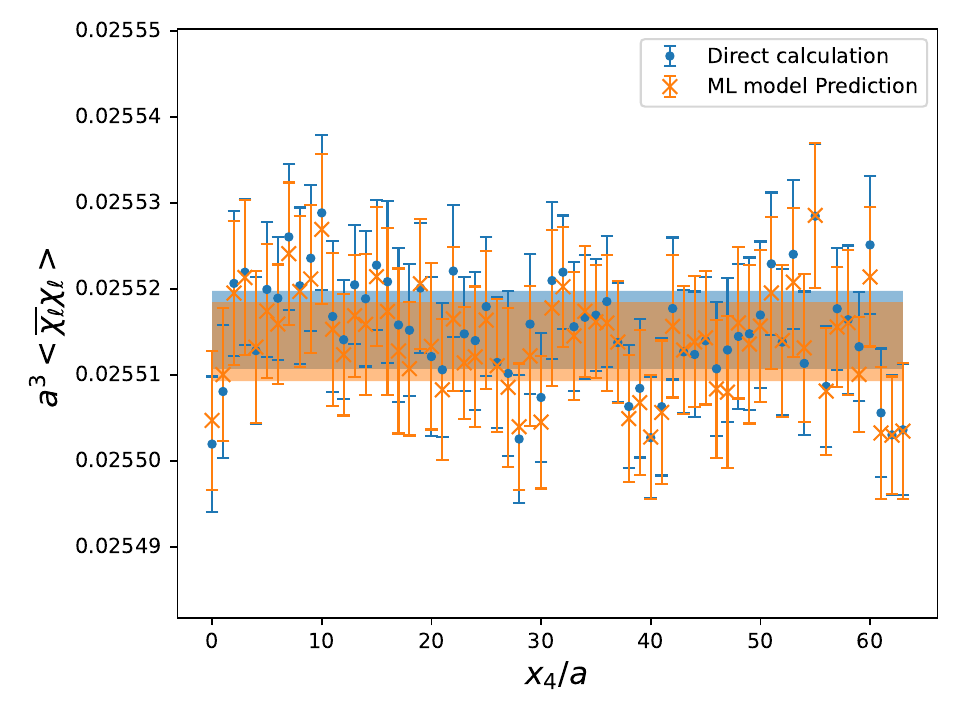}
  }
  \subfigure[With bias correction]{
\label{fig:pred_M1_flowtime_b}
  \includegraphics[width=0.3\textwidth]{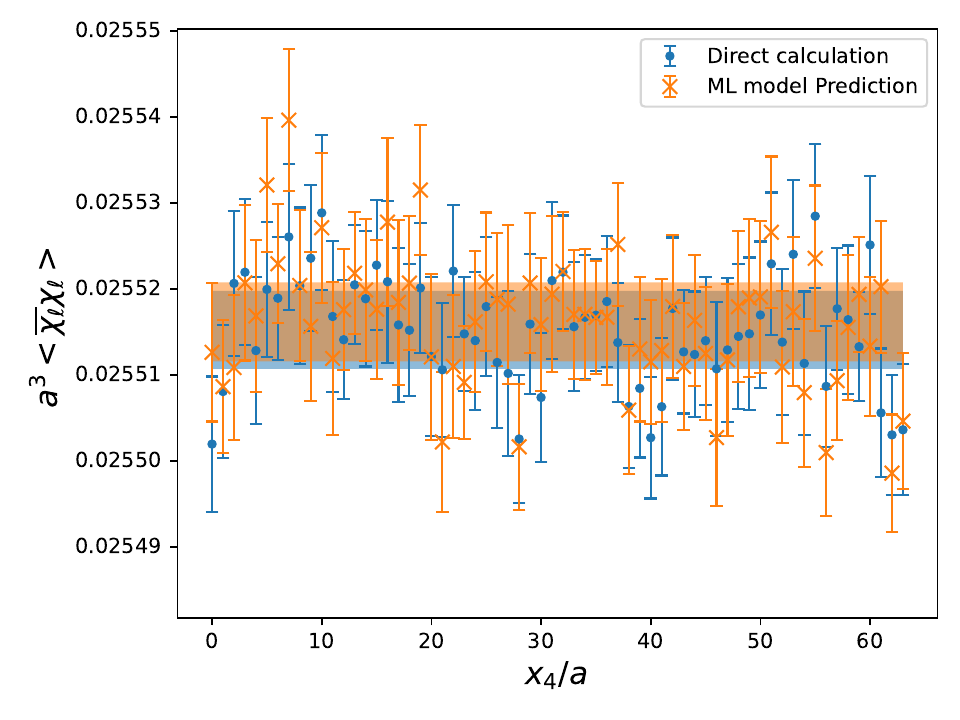}
  }
  \subfigure[Deviation between the direct and the ML calculation 
  normalized with the total standard deviation (see main text)]{
\label{fig:pred_M1_f}
  \includegraphics[width=0.3\textwidth]{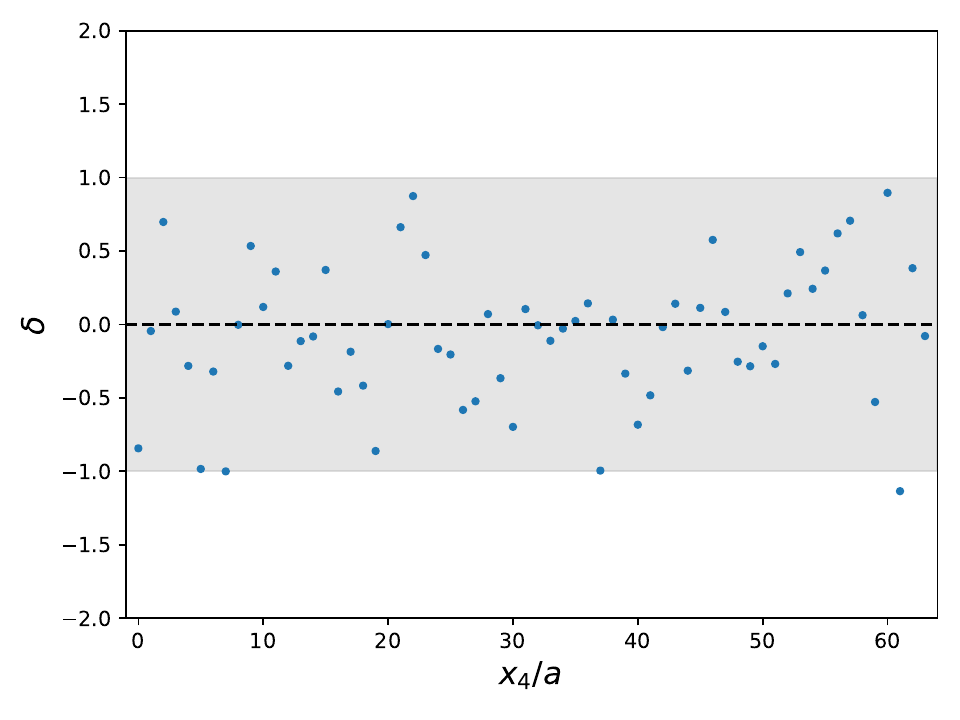}
  }
  \caption{ \label{fig:pred_M1_flowtime} 
  Comparison of the light quark condensate on the ensemble $M_1$ at flow time $t/a^2 = 0.7$ 
  between the ML result and the direct (standard) calculation of the same condensate as a function of Euclidean time. 
  The comparison is performed on the unlabeled data.
  }
\end{figure}

\begin{figure}[t]
	\subfigure[Without bias correction]{
\label{fig:pred_M3_flavor_a}
  \includegraphics[width=0.3\textwidth]{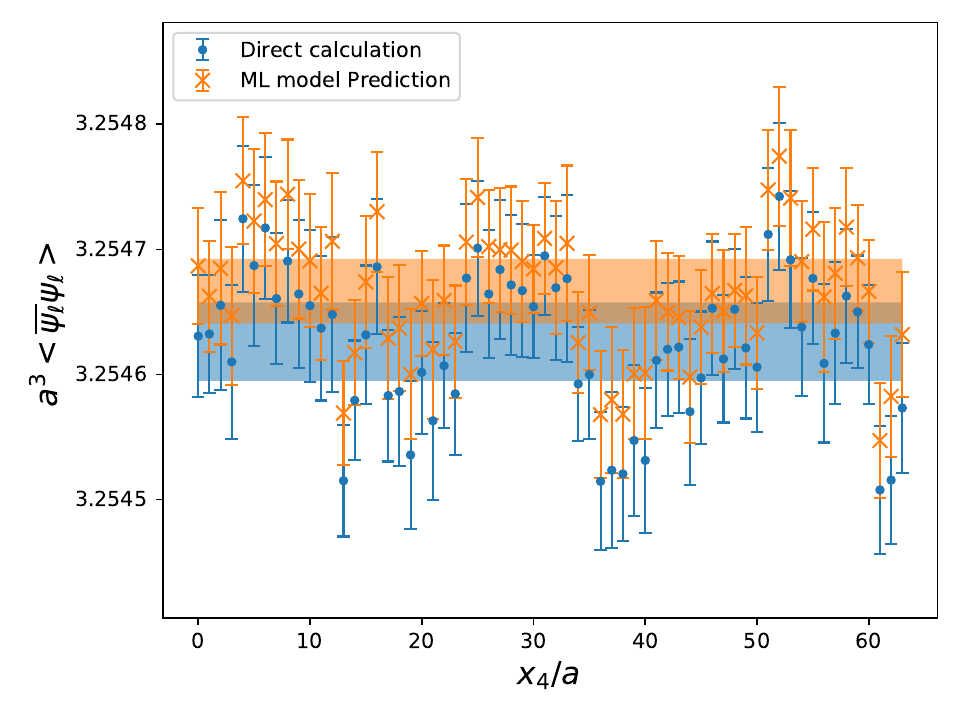}
}
	\subfigure[With bias correction]{
\label{fig:pred_M3_flavor_b}
	\includegraphics[width=0.3\textwidth]{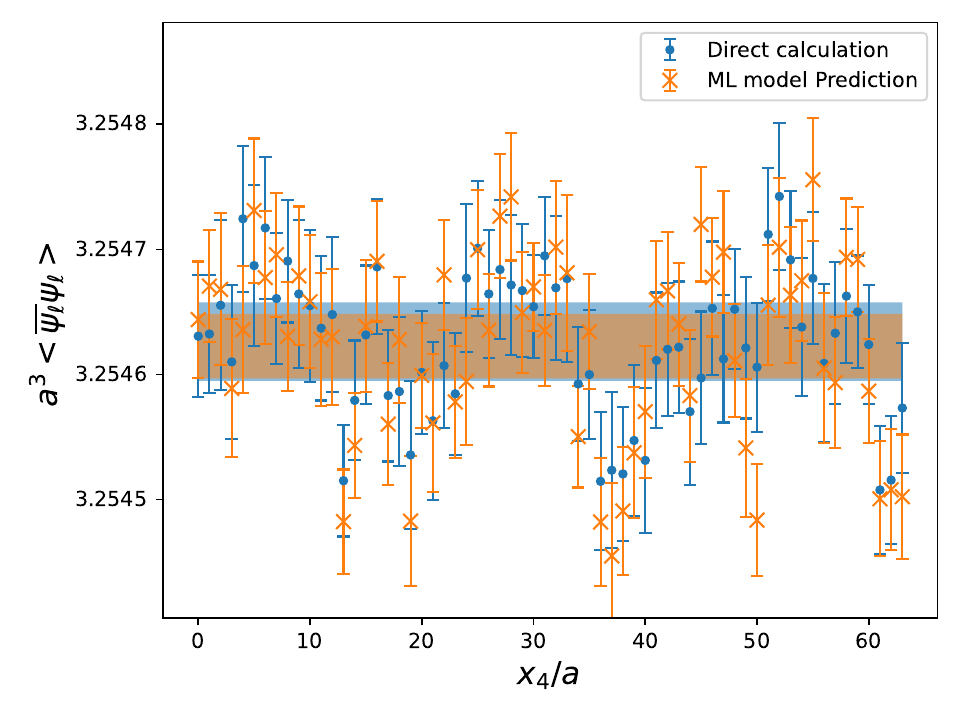}
}
  \subfigure[Deviation between the direct and the ML calculation 
  normalized with the total standard deviation (see main text)]{
\label{fig:pred_M3_q}
  \includegraphics[width=0.3\textwidth]{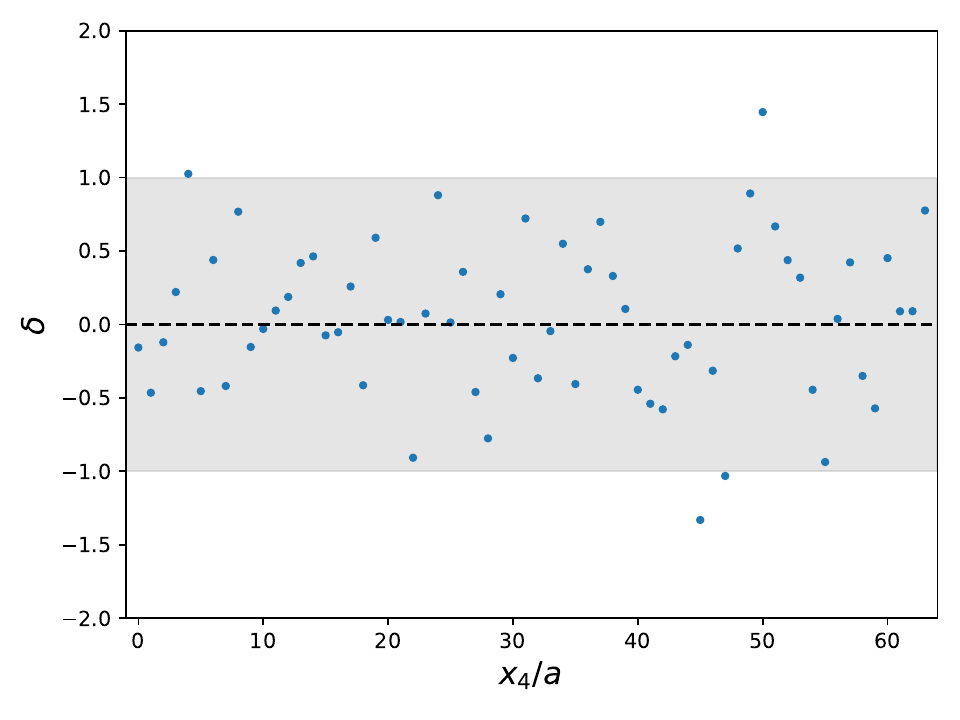}
}
  \caption{ \label{fig:pred_M3_flavor} 
  Comparison of the light quark condensate on the ensemble $M_3$ 
  between the ML result and the direct (standard) calculation of the same condensate as a function of Euclidean time. 
  The comparison is performed on the unlabeled data.
}
\end{figure}

Results after the training are shown in 
Figs.~(\ref{fig:pred_M1_flavor} - \ref{fig:pred_M3_flavor}),
where the ML results (orange data) 
are compared with a standard determination of the same condensates
on the unlabeled data (blue data).
With this choice both the standard and the ML determinations use 
a common set of data simplifying the comparison of statistical errors.
Figs.~\ref{fig:pred_M1_flavor} and \ref{fig:pred_M3_flavor} show the 
results for the light quark condensate at $t/a^2=0$ for the $2$ ensembles 
$M_1$ and $M_3$. The ML mappings are obtained using the strange quark condensate 
on the same ensembles.
Fig.~\ref{fig:pred_M1_flowtime} shows the result for the light quark condensate 
at $t/a^2=0.7$ obtained with a ML mapping trained with features given by 
the same condensate at $\bar{t}/a^2=0.5$.


For these $3$ cases, in the left plots we show the comparison between 
the ML and the direct determinations
before any bias correction is applied, while the middle plots are obtained after bias correction.
The right plots show the deviation between the direct and the ML calculation with bias correction, normalized by the total standard deviation.
\be 
\delta = \frac{\left\langle O \right\rangle (x_4) - \left\langle O_\Gamma \right\rangle (x_4)}{\sqrt{\sigma^2_{O_\Gamma} + \sigma^2_{O}}}\,.
\ee 
The statistical errors are calculated as follows.
First, we apply the ML mapping to the unlabeled data, after averaging over 
the stochastic sources, then we build blocks of $10$ elements and 
a standard bootstrap procedure is performed with $N_{\text{boot}}=1000$ bootstrap samples.
In this way, we determine the error bars of both the condensate for each $x_4/a$,
and the average over Euclidean time.

Comparing the plots before and after bias corrections we note that for both analysis for the ensemble $M_1$ the bias correction does not have any impact, because the ML prediction and the direct determination already agree perfectly
before applying the bias correction. In the case of the ensemble $M_3$ in Fig.~\ref{fig:pred_M3_flavor} we observe a slight bias of order of $1$ $\sigma$
that is corrected after applying a bias correction. In order to properly understand the dependence of the bias on the parameters of the data would require a much larger numerical study beyond the scope of this work.
Comparing the difference of the strange and light quark masses in the ensemble $M_1$ and $M_3$ (cfr. Table~\ref{tab:ensembles}) seems to suggest that at fixed labeled sets the larger mass difference in the ensemble $M_3$ implies a larger bias. 

\begin{figure}[t]
	\subfigure[Light quark condensate for the ensemble $M_1$ at $t/a^2=0$]{
\label{fig:pred_M1_flavor_nobias}
  \includegraphics[width=0.3\textwidth]{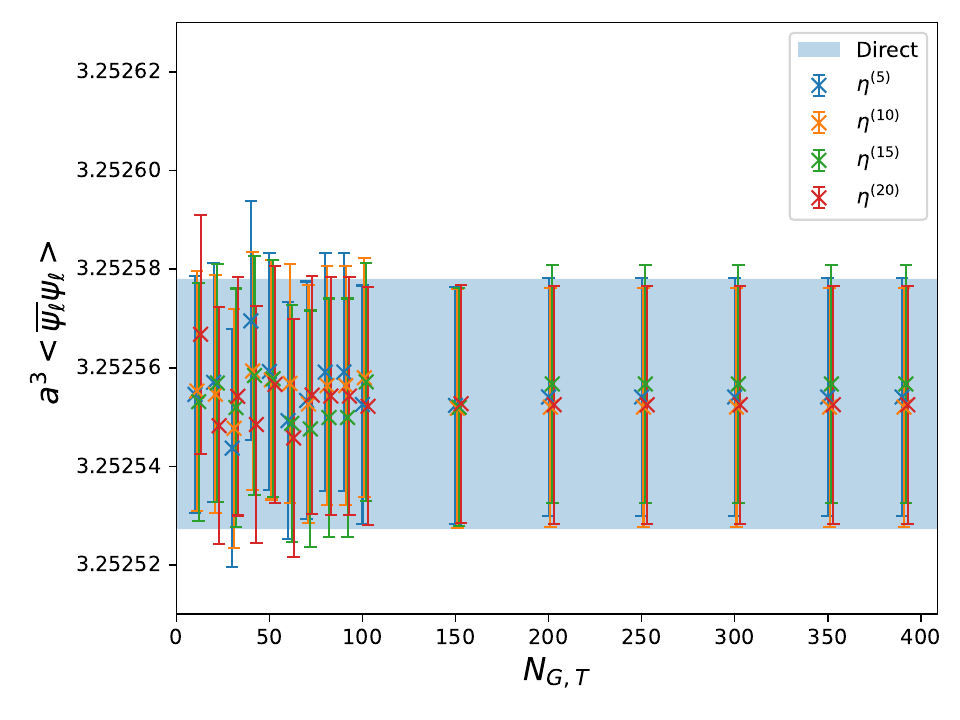}}
	\subfigure[Light quark condensate for the ensemble $M_1$ at $t/a^2=0.7$]{
\label{fig:pred_M1_flow_nobias}
	\includegraphics[width=0.3\textwidth]{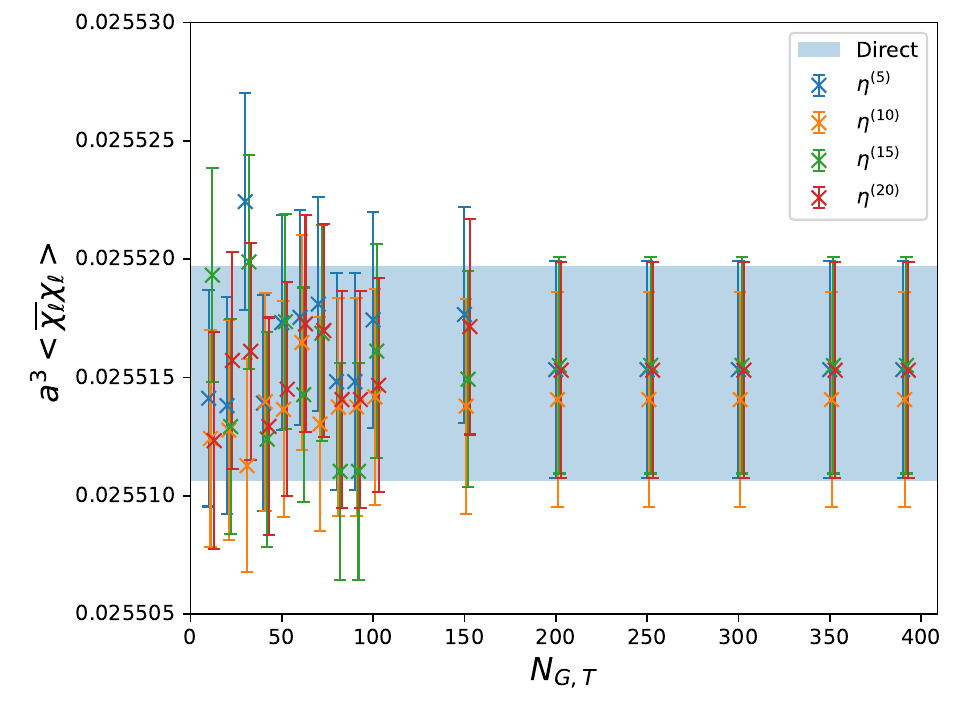}
  }
  \subfigure[Light quark condensate for the ensemble $M_3$ at $t/a^2=0$]{
\label{fig:pred_M3_flavor_nobias}
  \includegraphics[width=0.3\textwidth]{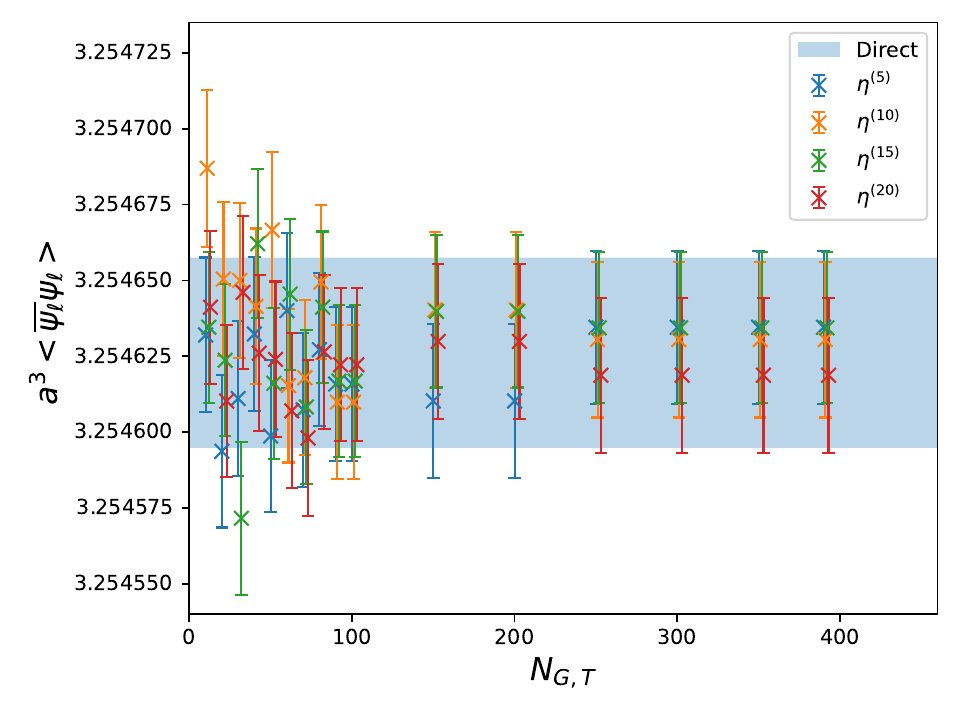}
  }
  \caption{Condensates determined directly on the unlabeled data (blue band) compared with the ML predictions as a function of the size of the training set $N_{G,T}$ for different source choices $\eta^{(r)}$. No bias correction is applied to the ML results.\label{fig:pred_nobias}} 
\end{figure}

To complement this observation, we have examined the impact of the training set size on the ML predictions prior to bias corrections, and the results are shown in Fig.~\ref{fig:pred_nobias}.
In Figs.~\ref{fig:pred_M1_flavor_nobias} and \ref{fig:pred_M1_flow_nobias}, 
showing the results for the ensemble $M_1$, we observe that increasing the size of the training set for $N_{G,T}>50$ provides no advantage consistently with the fact that for $N_{G,T}=50$ we observe no bias. For the ensemble $M_3$, with the results shown in Fig.~\ref{fig:pred_M3_flavor_nobias}, we observe a slight improvement increasing the size of the training set, but already for $N_{G,T}=100$ we observe no significant deviation between the direct and ML calculation. This analysis confirms that 
the bias correction is necessary only for the ensemble $M_3$. 
Given the cost-effectiveness of the training process,
while it is computationally efficient to increase the size of the training set for the ensemble $M_3$ to remove bias, we opted for a more conservative approach by calculating the bias in all ML determinations.

\begin{figure}[t]
	\centering
  \includegraphics[width=0.70\textwidth]{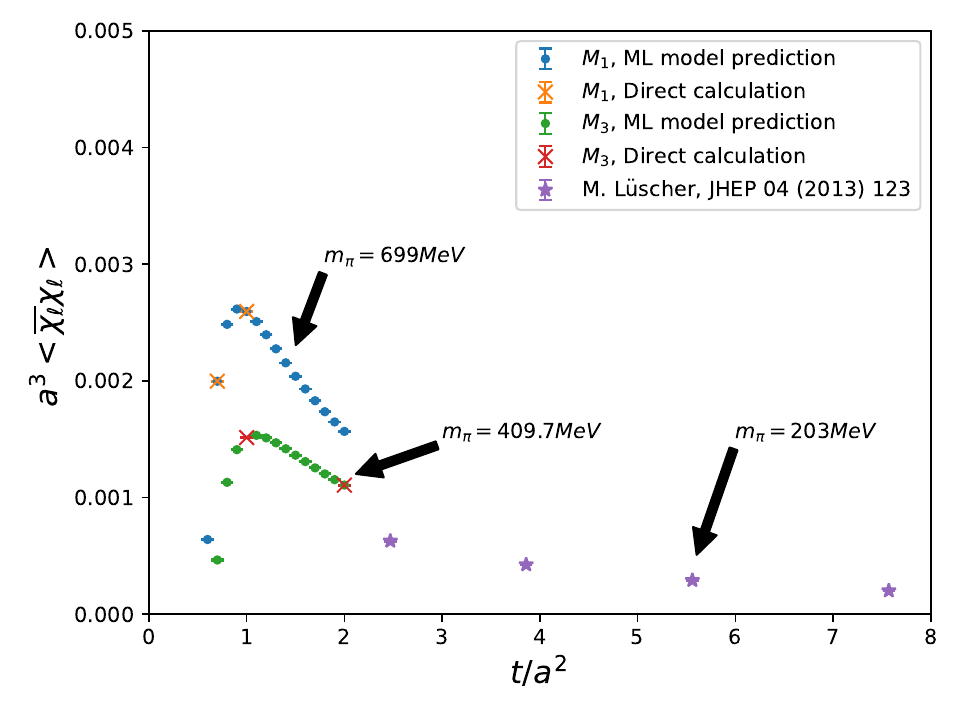}
  \caption{Flow time dependence of the light quark scalar condensate. The plot shows the results obtained with 
  a ML mapping and the results obtained in Ref.~\cite{Luscher:2013cpa}. With orange crosses 
  we also indicate flow times where we have available a direct calculation.
	}
  \label{fig:flow_cond}
\end{figure}

Using the trained DT we have computed the light quark condensate as a function of flow time across the entire range $0 \le t/a^2 \le 2$ for the ensembles $M_1$ and $M_3$.
The features employed to train the ML mapping consist of the light quark condensate at $\bar{t}/a^2=0.5$ for the ensemble $M_1$ and $\bar{t}/a^2=1.0$ for the ensemble $M_3$.
In Fig.~\ref{fig:flow_cond}, we present the ML results showing the 
light quark condensate as a function of flow time $t/a^2$.
To validate our result we show the results of a direct calculation at $t/a^2=0.7,~1.0$ for the ensemble $M_1$ and at $t/a^2=2.0$ for the ensemble $M_3$.
For comparison we also show results obtained at $t/a^2=1.0$ for the ensemble $M_3$.
The only calculation of the same quantity available in the literature is the result obtained in Ref.~\cite{Luscher:2013cpa}
with a direct calculation of the disconnected diagrams for larger values of the flow time. While our calculation does not cover the same range of flow times it gives us a qualitative comparison. The increase at small flow time for our data 
is easily explained.
The small flow time region covered by the ML calculation shows the power divergent contribution proportional to $m/t$ and possibly other power divergent contributions vanishing in the continuum limit.
In the chiral and continuum limit Ward identities connect the flowed condensate
with the physical condensate~\cite{Luscher:2013cpa,Shindler:2013bia},
and the power divergence in $1/t$ vanishes.
In Fig.~\ref{fig:flow_cond} we observe, at short flow time, the $1/t$
contribution that is suppressed at lower quark masses. For larger flow times 
and smaller quark masses, as described in Ref.~\cite{Luscher:2013cpa},
the $1/t$ contribution is small and the remaining logarithmic contribution 
is cancelled by the contribution of the vacuum-to-pion 
pseudoscalar matrix element evaluated at the same flow time.

\subsection{Robustness of the ML mapping}
\label{ssec:robust}

To test the robustness of the DT we trained we have varied 
specific hyperparameters, like the number and the choice of 
gauge configurations and stochastic sources used for the training.
We have studied the dependence on the choice of the training set 
considering different, randomly chosen,
sets of $N_{G,T} = 50$ gauge configurations 
and different choices for the stochastic source.

\begin{figure}[t]
  \centering
   \subfigure[$\left\langle \psibar_\ell \psi_\ell \right\rangle$ from ensemble $M_1$]{
  \includegraphics[width=0.31\textwidth]{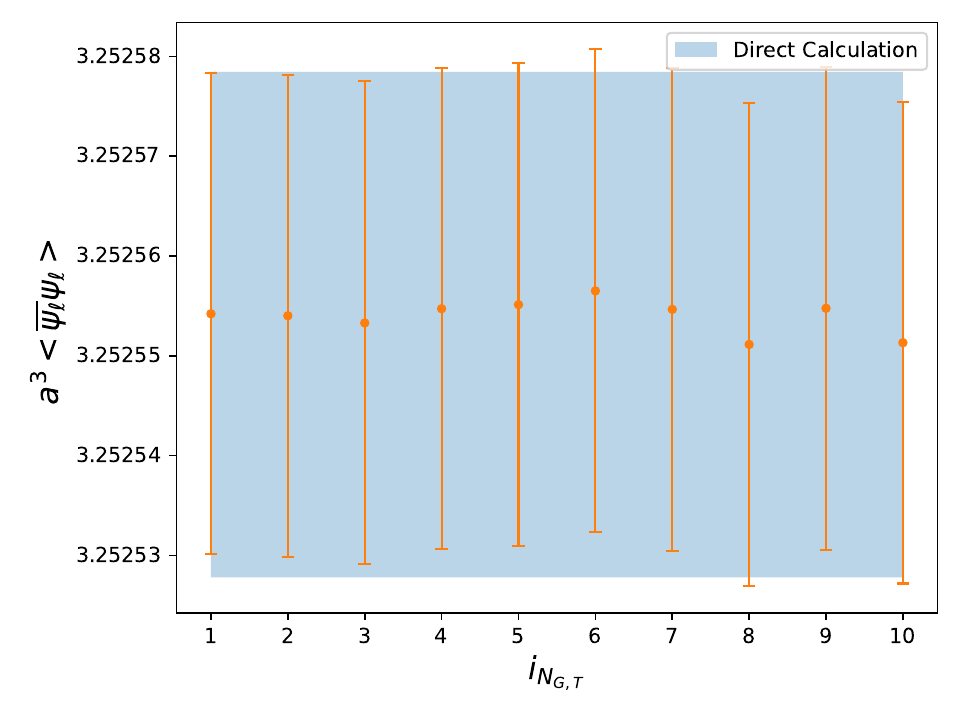}
   }
   \subfigure[$\left\langle \chibar_\ell \chi_\ell \right\rangle$ at $t/a^2 = 0.7$ from ensemble $M_1$. The ML mappings have been trained at $\bar{t}/a^2=0.5$.]{
         \includegraphics[width=0.31\textwidth]{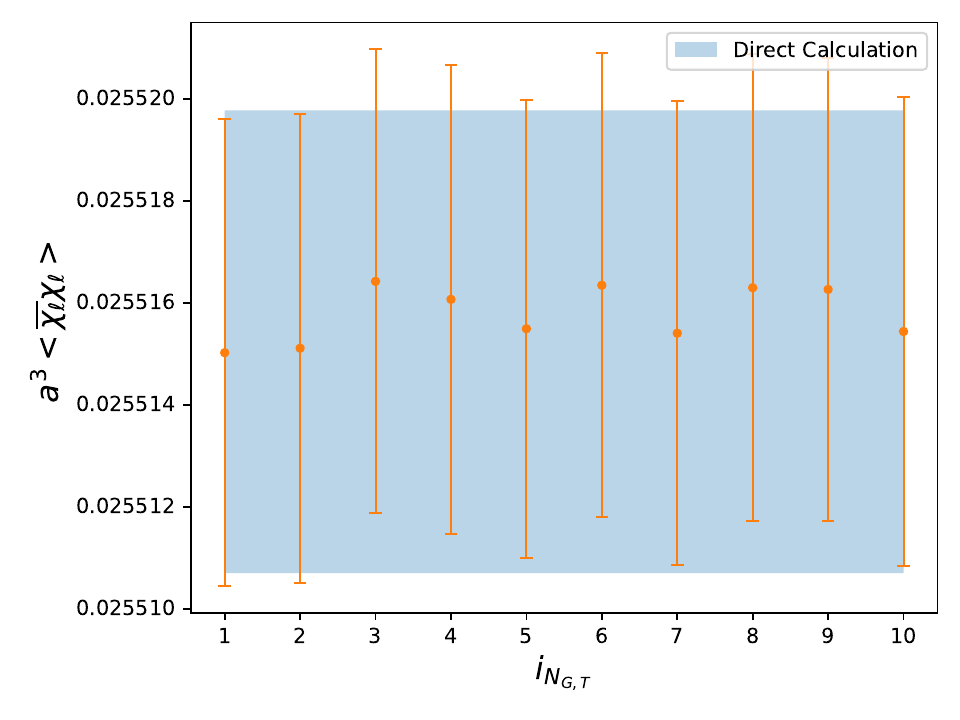}
   }
   \subfigure[$\left\langle \psibar_\ell \psi_\ell \right\rangle$ from ensemble $M_3$]{
         \includegraphics[width=0.31\textwidth]{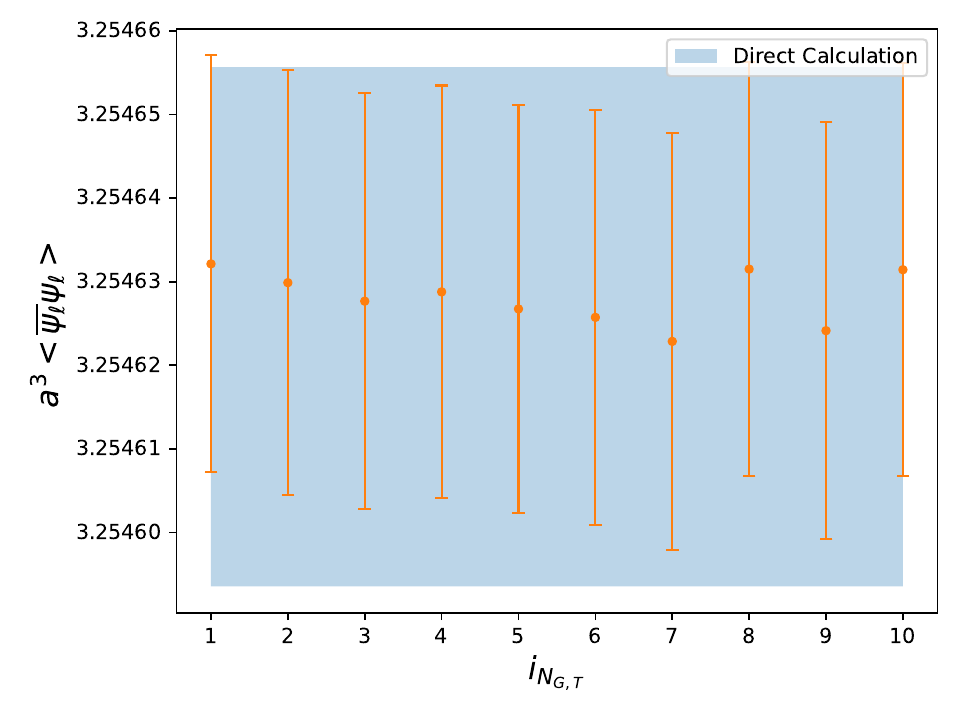}
   }
 \caption{
	 \label{fig:Ntrain_dep}
       Results for the light quark condensates for $10$ different training sets, labeled by $i_{N_{G,T}}$,
       obtained selecting randomly $N_{G,T} = 50$ gauge configurations among the 
       full ensemble. The blue band represents the calculation 
			 of the light quark condensate on the full set of unlabeled data $N_{U}$.
   }
\end{figure}

\begin{figure}[t]
  \centering
 \subfigure[$\left\langle \psibar_\ell \psi_\ell \right\rangle$ from ensemble $M_1$]{
\label{fig:src_dep_M1_flavor}
 \includegraphics[width=0.3\textwidth]{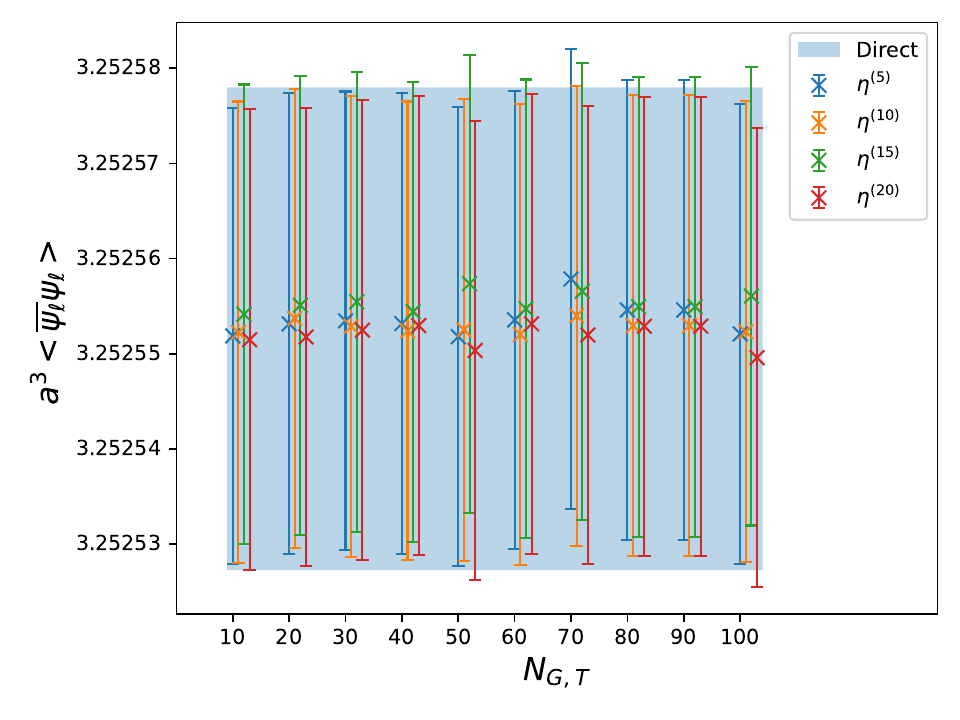}
 }
 \subfigure[$\left\langle \chibar_\ell \chi_\ell \right\rangle$ at $t/a^2 = 0.7$ from ensemble $M_1$. The ML mappings have been trained at $\bar{t}/a^2=0.5$.]{
\label{fig:src_dep_M1_flowtime}
 \includegraphics[width=0.3\textwidth]{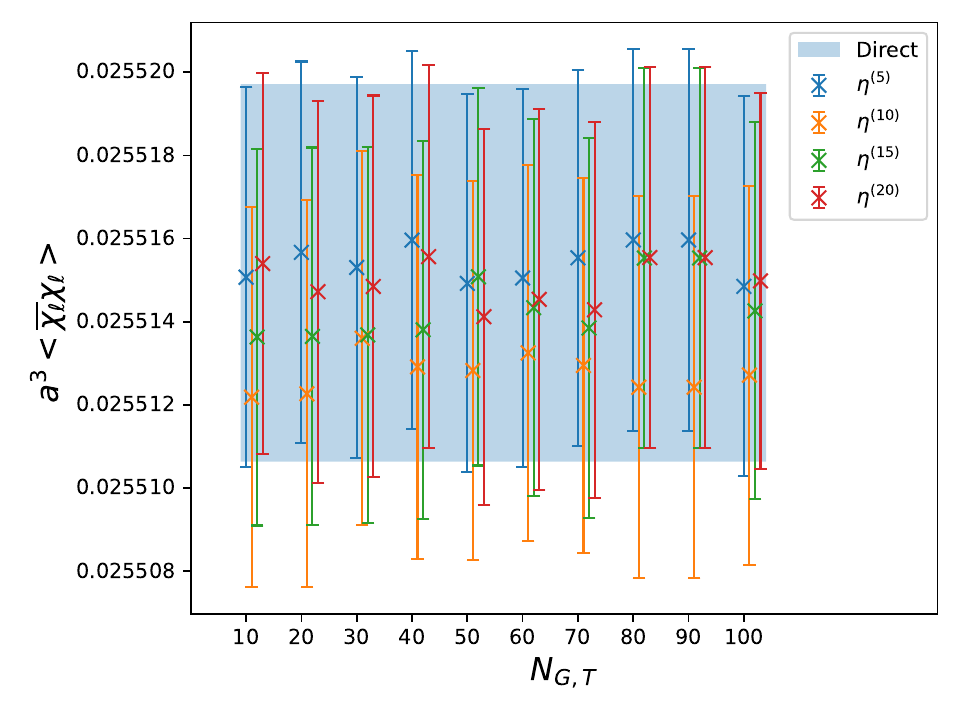}
 }
 \subfigure[$\left\langle \psibar_\ell \psi_\ell \right\rangle$ from ensemble $M_3$]{
\label{fig:src_dep_M3_flavor}
 \includegraphics[width=0.3\textwidth]{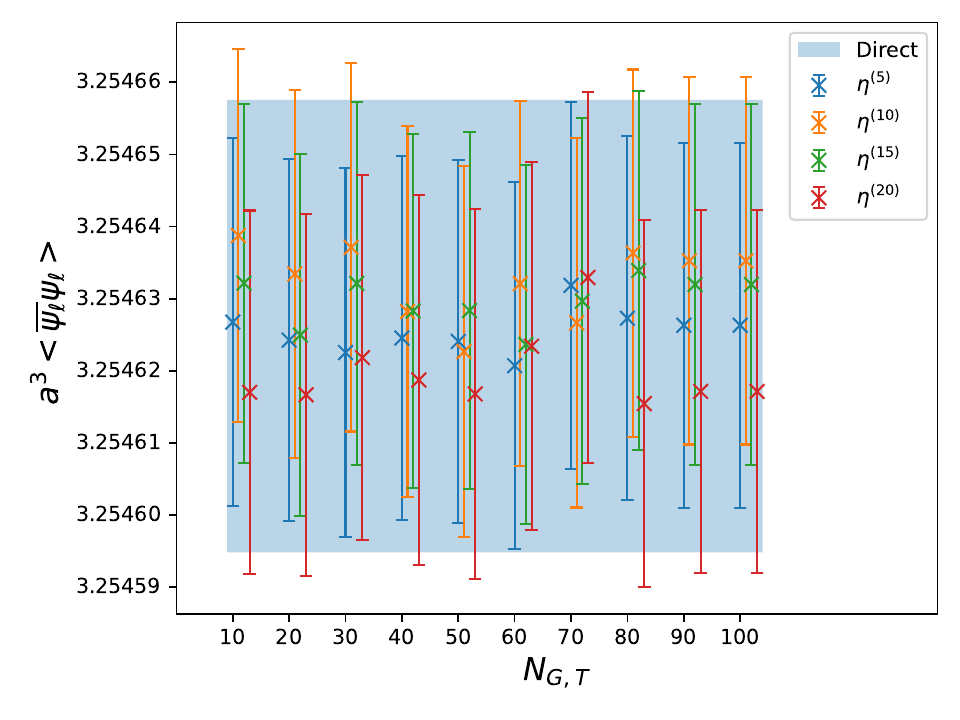}
}
 \caption{ \label{fig:bias_corr_src_dep} Dependence of the quark condensates
 on the choice of stochastic sources and size of the training set, compared with the direct calculation.
The direct calculation, shown with the blue band, utilizes the full set of unlabeled data $N_{U}$. The ML mapping calculation utilizes in the training set different single stochastic sources, $\eta^{(r)}$, $r=5,~10,~15,~20$, and different numbers of gauge configurations, $N_{G,T}$. 
The bias correction is calculated with $N_{G,B} = N_G - N_{G,T}$ gauge configurations and the same stochastic source used for the training.
	}
\end{figure}

In Fig.~\ref{fig:Ntrain_dep} we show the ML calculation 
for $10$ different random choices of the $N_{G,T} = 50$ gauge configurations
used for the training, 
and compare it with the full direct calculation of the complete set, $N_U$, of unlabeled data.
We hardly observe any deviation between all the calculations.
In Fig.~\ref{fig:bias_corr_src_dep} we show the dependence 
of the ML calculations 
on the choice of the stochastic source and on the 
number of gauge configurations, $N_{G,T}$, used for the training.
We observe no significant deviation when we choose a different stochastic 
source or when we vary the size of the training set.
We conclude that changing the training set or its size results 
in little to no deviation between the different ML mappings obtained
and the direct standard calculation.

\section{Conclusions}
\label{sec:conclusions}

We have investigated an application of 
supervised Machine Learning (ML) techniques for lattice QCD calculations.
Fermionic disconnected diagrams are among the most expensive
quantities to calculate in any lattice QCD computation.
We have trained a ML mapping 
to speed up the calculation of fermionic disconnected diagrams 
for a set of external parameters such as the quark mass and the flow time.
The mapping is a decision tree trained with a subset of the 
full set of data usually analyzed for a standard lattice QCD calculation.
After applying bias correction, we find that the condensates 
calculated with ML deviate at most by $1$ sigma over the whole 
set of parameters investigated, while maintaining consistent statistical uncertainties.

The computational gain depends on whether the ML mapping is trained using $2$
different quark masses or $2$ different flow times.
In the first case, given the time needed for a standard lattice QCD calculation of the light condensate on $N_U$ data, the gain applying a ML mapping is given by 
\be 
\text{Gain}= \left(\frac{t_s}{t_d}\frac{N}{N_U}+ \frac{N_L}{N_U}\right) \,,
\ee 
where $t_{d,s}$ label the time needed to calculate the down, or strange, 
condensate for a single 
stochastic source and single gauge configuration.
We have estimated the timings for the calculation of the quark propagators,
and conclude that the ML calculation requires $72-76\%$ of the time needed for 
a standard computation. The gain depends almost solely on the 
difference in computer time needed for the calculation of the $2$ quark propagators used in the ML method. We have compiled the gains achieved in the two ensembles we have analyzed in Table~\ref{tab:gain}.
The gains are expected to increase further for lighter pion ensembles,
owing to the larger difference between the strange and quark masses.
However, it remains to be investigated whether there is sufficient data correlation to effectively apply this ML method in such cases.

\begin{table}[h]
  \centering
  \begin{tabular}{|c|c|c|c|c|c|c|c|c|}
    \hline
    Ens. & Input & Obs. & Direct  & ML & $N_T$ & $N_B$ & $N_U$ & Gain \\
    \hline
    \hline 
		M$_1$ & $a^3 \left.\left\langle \psibar \psi \right\rangle\right|_{m_s}$ &  $a^3 \left.\left\langle \psibar \psi \right\rangle\right|_{m_\ell}$ & 3.252553(25) & 3.252554(24) & 50 & 349 & $399 \times 19 $ & 1.31 \\
    \hline 
		M$_3$ & $a^3 \left.\left\langle \psibar \psi \right\rangle\right|_{m_s}$ &  $a^3\left.\left\langle \psibar \psi \right\rangle\right|_{m_\ell}$ & 3.254626(31) & 3.254622(26) & 50 & 400 & $450 \times 19 $ & 1.39 \\
    \hline
    \hline
		M$_1$ &  $a^3\left.\left\langle \chibar \chi \right\rangle\right|_{t=0.5}$ &  $a^3\left.\left\langle \chibar \chi \right\rangle\right|_{t=0.7}$ & 0.0019974(48) & 0.0019970(47) & 50 & 349 & $399 \times 19$& 1.26 \\
    \hline                          
		M$_1$ & $a^3 \left.\left\langle \chibar \chi \right\rangle\right|_{t=0.5}$ &  $a^3 \left.\left\langle \chibar \chi \right\rangle\right|_{t=1.0}$ & 0.0025947(35) & 0.0025946(34) & 50 & 349 & $399 \times 19$ & 1.71 \\
		\hline                                                                      M$_3$ & $a^3 \left.\left\langle \chibar \chi \right\rangle\right|_{t=1.0}$ & $a^3 \left.\left\langle \chibar \chi \right\rangle\right|_{t=2.0}$ & 0.0011067(37) & 0.0011070(33) & 50 & 400 & $450 \times 29$ & 1.80 \\
    \hline
  \end{tabular}
  \caption{Selected results comparing different different observables predicted and adopted to train the ML mapping. The corresponding training, $N_T$, bias, $N_B$, and unlabeled, $N_U$, set are shown together with the computational gain (see main text)
    \label{tab:gain}}
\end{table}

\begin{figure}[h]
  \centering
 {
\label{fig:gain_single}
 \includegraphics[width=0.45\textwidth]{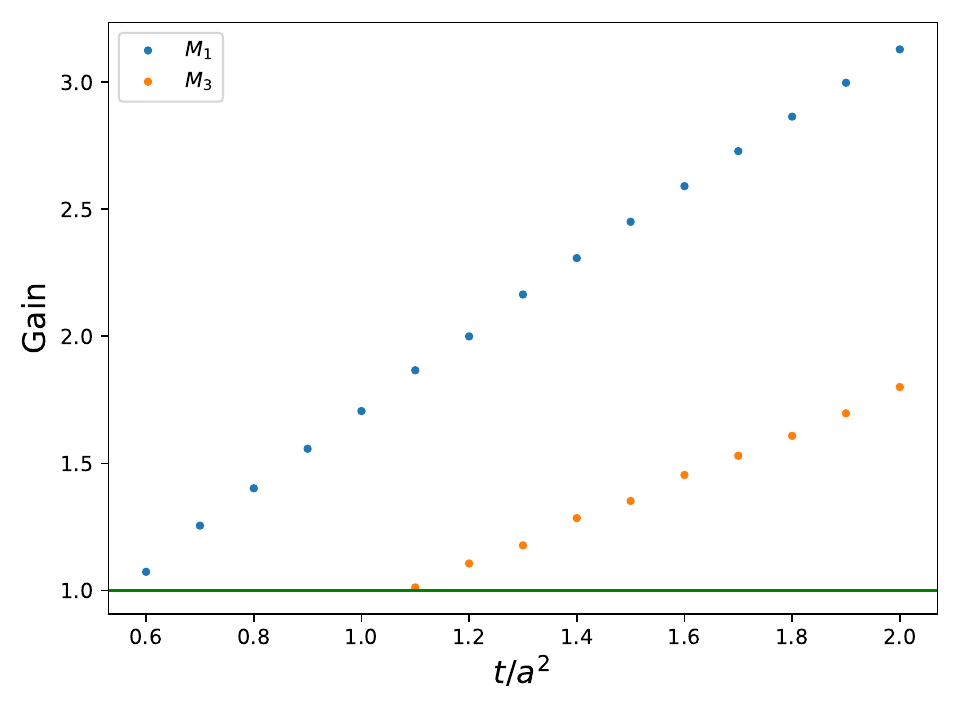}
 }
  {
\label{fig:gain_summed}
 \includegraphics[width=0.45\textwidth]{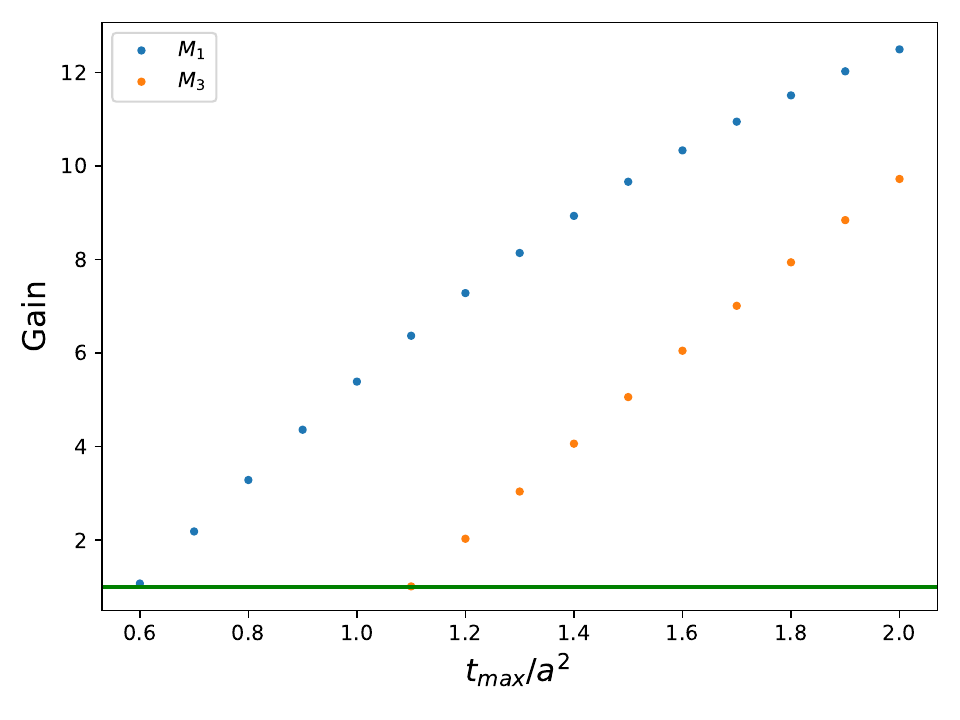}
 }
 \caption{\label{fig:gain}
 Left plot: Gain obtained as function of the flow time. To 
 train the ML mapping we use the quark condensate at $\bar{t}/a^2=0.5$ for the ensemble $M_1$ and $\bar{t}/a^2=1.0$ for the ensemble $M_3$. Right plot: Gain obtained as function of the maximal flow time, $t_{\text{\tiny{max}}}/a^2$. To 
 train the ML mapping we use the quark condensate at $\bar{t}/a^2=0.5$ for the ensemble $M_1$ and $\bar{t}/a^2=1.0$ for the ensemble $M_3$. We calculate the light quark condensates using ML for all the flow times, from $t/a^2=0.5$ for the ensemble $M_1$ and $t/a^2=1.0$ for the ensemble $M_3$, at interval of $0.1$ up to $t_{\text{\tiny{max}}}/a^2$.
  }
\end{figure}

In the case of different flow times the gain depends on the number of flow times $N_t$ where the condensate is determined and on the different flow times used in the training procedure. In the case we use the ML mapping to compute the chiral condensate at a single flow time, 
in Table~\ref{tab:gain} we have collected the gains achieved 
in three different cases. The gains are also shown in the left plot 
of~Fig.~\ref{fig:gain} as a function of the flow time $t/a^2$ of the calculated condensate. For this plot the features are given by the 
light quark condensate at flow time $\bar{t}/a^2=0.5$ for the ensemble $M_1$ and $\bar{t}/a^2=1.0$ for the ensemble $M_3$. We observe an almost linear gain as a function of the difference between the input and output flow time up to a factor $3$ for the ensemble $M_1$ at flow time $t/a^2=2.0$ and a factor $1.8$ for the ensemble $M_3$. We do not show any data for $t<\bar{t}$ because 
there is no gain in trying to apply ML to compute the condensate for flow times smaller than the input feature. 

If we train the ML mapping using the same features ($\bar{t}/a^2=0.5$ for the ensemble $M_1$ and $\bar{t}/a^2=1.0$ for the ensemble $M_3$) for many flow times $t>\bar{t}$ we obtain a larger gain.
In the right plot of Fig.~\ref{fig:gain} we show the gain if we train the ML for flow times at intervals of $0.1$ for $t>\bar{t}$ as a function of the maximal flow times utilized, $t_{\text{\tiny{max}}}/a^2$. 
In this case we obtain gains of a factor $10$ for $M_3$ and $12$ for $M_1$.

Machine Learning methods can provide powerful computational tools to make 
better use of the plethora of data produced in standard lattice QCD calculations.
In this work we have successfully applied a supervised ML method to speed up the calculation of disconnected fermionic diagrams.
We consider this work as a first attempt into the exploration of novel 
paths for the determination of the quark propagator and fermionic correlation 
functions in lattice QCD simulations.

\section*{Acknowledgments}
We thank Tom Luu for constant encouragement and a critical reading of the manuscript.
We have profited from discussions with A. Bazavov.
We acknowledge the Center for Scientific Computing, University of
Frankfurt for making their High Performance Computing facilities available.
The authors also gratefully acknowledge the computing time granted by the 
JARA Vergabegremium and provided on the JARA Partition part of the 
supercomputer JURECA~\cite{jureca:2021} at Forschungszentrum Jülich.
J.K. was supported by the Deutsche Forschungsgemeinschaft (DFG, German Research Foundation) 
through the funds provided to the Sino-German Collaborative Research Center TRR110 
"Symmetries and the Emergence of Structure in QCD" (DFG Project-ID 196253076 - TRR 110).
A.S. acknowledges funding support from Deutsche Forschungsgemeinschaft 
(DFG, German Research Foundation) through grant 513989149 and 
under the National Science Foundation grant PHY- 2209185.
G.P. is funded by the Deutsche Forschungsgemeinschaft (DFG, German Research Foundation) - project number 460248186 (PUNCH4NFDI).
We acknowledge support from the DOE Topical Collaboration “Nuclear Theory for New Physics”, award No. DE-SC0023663.

\appendix
\section{Hyperparameters of the model}
\label{app:hyperparams}

For the decision tree regression adopted in this work we have used the module 
{\tt sklearn.tree} in the scikit-learn library.
In particular we adopt the {\tt DecisionTreeRegressor} class that allows to create, train, and use decision tree models for regression problems.
We list here the choices of hyperparameters made in this study.
The function to measure the quality of a split is the mean squared error (MSE), i.e. we minimize the minimization the L2 loss using the mean of each terminal node. This is equivalent to variance reduction, as feature selection criterion. A node will be split if this split induces a decrease of the MSE greater than zero.
The nodes are expanded until all leaves are pure or until all leaves contain less than $1$ sample. This means that a split point at any depth will only be considered if it leaves at least $1$ training samples in each of the left and right branches. This may have the effect of smoothing the model, especially in regression.
When looking for the number of features to consider for the best split we consider their total number and permute them randomly at each split. We do not include any limit in the number of leaf nodes and we do not perform any pruning.

\bibliography{ml-paper}
\end{document}